\documentclass[12pt]{article}
\usepackage{epsfig, bm, amsfonts, amssymb}
\usepackage{multirow}
\usepackage{tabularx}
\usepackage{hhline}
\usepackage[normalem]{ulem}
\usepackage{cite}
\usepackage{color,soul}
\usepackage[none]{hyphenat}
\usepackage[textsize=tiny]{todonotes}
\usepackage[12pt]{moresize}
\usepackage{chngcntr}
\usepackage[export]{adjustbox}
\usepackage{url}
\usepackage[T1]{fontenc}
\usepackage[icelandic,english]{babel}

\counterwithin{figure}{section}
\counterwithin{table}{section}
\counterwithin{equation}{section}

\RequirePackage{color}

\newcommand{\blue}[1]{{\color{red}{#1}}}

 \definecolor{MyDarkGreen}{rgb}{0.02,0.60,0.06}

\def\CGG{\emph{Cogadh Gaedhel re Gallaibh}}
 \def\CG{\emph{Cogadh}}

\def\NS{\emph{Nj{\'{a}}ls Saga}}
\def\NM{N{\'{\i}} Mhaonaigh}
\def\DC{{D{\'{a}}l Cais}}

\title{{\bf Network Analysis of \\the Viking Age in Ireland \\ as portrayed in \\ {\CGG}}}
\author{ 
 {\it Joseph Yose}$^{1,*}$,
 {\it Ralph Kenna}$^{1,*}$,\\
 {\it  M{\'{a}}ir{\'{\i}}n MacCarron}$^2$
 and
  {\it P\'{a}draig MacCarron}$^3$ \\~\\
	\begin{footnotesize}
\begin{tabular}{ll}
  $1$& {{Applied Mathematics Research Centre, Coventry University, Coventry CV1 5FB, England}}\\
  $2$& {{Department of History, University of Sheffield, Sheffield S3 7RA, England}} \\ 
	$3$& {{Social \& Evolutionary Neuroscience Research Group, Department of Experimental}}\\
	    & {{Psychology, University of Oxford, Oxford OX1 3UD, England}}
\end{tabular}
	\end{footnotesize}
{}\\~\\
}

\begin{document}
\maketitle
\vspace{-1cm}
%-----------------------------------------------------------------------
{\Large
  \begin{abstract}
%-----------------------------------------------------------------------
%
\noindent

{\CGG}  (``The War of the Gaedhil with the Gaill'')
is a medieval Irish text, {telling} how an army under the leadership of Brian Boru challenged Viking invaders and their allies in Ireland, culminating with the Battle of Clontarf in 1014. Brian's victory is widely remembered for breaking Viking power in Ireland, although much modern scholarship {disputes}  traditional perceptions. Instead of an international conflict between Irish and Viking, {interpretations based on} revisionist scholarship consider it a domestic feud or civil war. {Counter-revisionists} challenge this view and a longstanding and lively debate continues. Here we {introduce} quantitative measures to the discussions. We present statistical analyses of network data embedded in the text to position its sets of interactions on a spectrum from the domestic to the international. This delivers a picture that lies between antipodal traditional and revisionist extremes; hostilities recorded in the text are mostly between Irish and Viking - but internal conflict forms a significant proportion of the negative interactions too.

\vspace*{\fill} 
\noindent
{\underline{~~~~~~~~~~~~~~~~~~~~~~~~~~~~~~~~~~~~~~~~~~~~~~}}\\
$^\ast {\mathbb L}^4$ Collaboration for the Statistical Physics of Complex Systems,   
Leipzig-Lorraine-Lviv-Coventry, Europe.

%-----------------------------------------------------------------------
                        \end{abstract} }
%-----------------------------------------------------------------------
%
  \thispagestyle{empty}
%
%***********************************************************************
%
  \newpage

%
%-----------------------------------------------------------------------
%                  \pagenumbering{arabic}
%-----------------------------------------------------------------------

%%%%%%%%%%%%%%%%%%%%%%%%%%%%%%%%%%%%%%%%%%%%%%%%%%%%%%%%%%%%%%%%%%%
\section{Introduction}
%%%%%%%%%%%%%%%%%%%%%%%%%%%%%%%%%%%%%%%%%%%%%%%%%%%%%%%%%%%%%%%%%%%
\label{sec1}

Modern academic disciplines do not exist in isolation and are increasingly interdependent and interconnected.
For example, our understanding of the past utilises scientific analyses of archaeological data,
anthropology derives from evolutionary biology and economics requires mathematics and statistics.
Statistical-physics inspired methodologies have long been applied to other academic disciplines, motivated not least by curiosity as to how complex systems emerge from interactions between constituent parts in non-trivial manners. 
Scientific curiosity of this kind has led to the development of new interdisciplinary areas 
and the creation of  new knowledge by thinking beyond traditional methodological boundaries.
In recent years, facilitated by new access to extensive data sets and technological progress, many statistical physicists have broadened their interests to include network science, a methodology which has led to an explosion of interdisciplinary activity.
While many social-network studies focus on modern forms of sociality such as online communications and other forms of computer-mediated social media, the importance of exploring other kinds of data is increasingly recognised as well.
In particular, quantitative investigations of epic narratives can advance our understanding of the past.
A plethora of quantitative approaches and suggestions to investigate societal and cultural aspects of the past 
%past societies and cultures 
are contained in the compendium \cite{MMM}. Here we apply and develop one such method to a long-standing debate about the Viking age in Ireland.

%\todo{Intro to CGG + debate}
The Battle of Clontarf (1014), an iconic event in the history of Ireland, is traditionally remembered as marking the decline of  Viking power after some two centuries in the country.
For the past 250 years a debate has been taking place centered around what may be called
``traditionalist'' and ``revisionist'' views of the period
\cite{OConnor1766,OHalloran,Todd,Ryan1938,Corrain72,Duffy}.
The  recent millennial anniversary of the Battle  inspired academics to revisit the debate through new journal papers, books, booklets, monographs, online commentaries and media engagements 
(e.g., Refs.\cite{Birkett,CaseyMeehan2014,Clarke,Duffyproc,Duffy,IrishTimes,Duffyreview2,McGettigan,RIA,UCD,NMCam,Downham2014}).
As with earlier investigations, these approaches treat the subject matter using traditional tools of the humanities 
(e.g., Refs.\cite{Bugge05,Bugge,Meyer10,Th21,Goedheer,RyanonGoedheer,Flower1947,Gh81,NMbias,Holm1994,NMdating,NMannals,Mhaonaigh97,Corrain97,Corrain98,NMfoe,Mhaonaigh02,Downham2005,Casey2010,DuBois,Urdail,NMneglect,Casey2013,MhaonaighTroy,Kirwan,Smyth,Downham2007,DownhamScot,Downham2012a,DownhamAnn}).
Here we present an alternative, complexity science-based investigation, using one of the 
most famous accounts of the Vikings in Ireland: {\CGG}\footnote{Alternative spellings exist in the 
						 literature but we employ the spelling  used by James Henthorn Todd \cite{Todd} since his is the edition that we analyse. 
						 We sometimes refer to the narrative simply as ``the {\CG}'' hereafter.} (``The war of the Gaedhil with the Gaill'' or ``War of the Irish with the Foreigners'').
						%, which \red{probably} dates from the late 11th or early 12\red{th} centuries.

%\margintext{90}{{Intro to CGG + debate}}

%\todo{The text and its network}
{The Viking age in Ireland approximately spans the ninth to twelfth centuries.}
The  {{\CG}} starts with the arrival of the Vikings\footnote{There 
          are a number of 
          etymological theories for the word ``Viking''~{\cite{Downham2012a}}. 
          We use it to refer to the  medieval Norse or Scandinavian  raiders and invaders 
					who attacked Ireland (and other countries) by sea,  or those who subsequently settled in Ireland, 
					between the late 8th  and 11th centuries \cite{Kirwan}. 
					A stricter definition of the term ``Viking'' may involve the notion of ``piratical'' and  
					in this sense, not all Vikings were Scandinavian and not all Scandinavians were Vikings \cite{Kirwan}.
					But we use the term in the looser sense (in keeping with much of the literature, e.g., 
					Refs.\cite{Casey2013,Downham2005,DuBois,Corrain98,NMneglect,NMbias,NMannals,NMfoe}).
These are the {\emph{Gaill}}  (singular {\emph{Gall}})  referred to above.
%\blue{Clare Downham discusses cohesiveness of Viking identities in Ref.\cite{Downham2012a} and the appearance of groups called Gallgo{\'{\i}}dil  or ``foreigner Gaels'' who may ``assimilate elements of their host culture and develop new hybrid identities.'' She goes on to discuss how ``a notion of Scandinavian heritage was consciously maintained among these groups supporting ties to other viking communities.’’}
}   (in 795) and gives a chronicle of their various raids.
%Todd page xxxiii says ``We may, therefore, safely' adopt the year 795, on the united authority of the Irish and W elBh Annals, as the real date of the first appearance of Scandinavian pirates in the Irish seas.''
%See alsopage xxxii where he talks about Recbru as Rathlinn island.
This is followed by a discussion of the Irish {\DC} dynasty, their deeds, and those of their leader, Brian Boru, culminating in the Battle of Clontarf in 1014.
% See Todd page xxviii) We may now proceed to give a more particular account
% to of the contents of the present work, which divides itself
%into two parts. The first part ends with the chapter
%numbered 1 XL., and contains an account in chronological
%order, or what is meant to be so, of the arrival of the
%"fleets" of the Norsemen in different parts of Ireland,
%especially the southern or Munster district. The second
%part, from chap. XLI. to the end, is devoted to the history
%of the Dal Cais, or Munster Chiefta.ins, and particularly
%to the achievements of their great hero, Brian, hiH 1l8lll'pation
%of the throne of Ireland, for such it was, and his
%death in the celebrated Battle of Clontarf.
%
%Wikipedia says: 
%CGG is a medieval Irish text that tells of ... 
%Irish king Brian Boru's great war against them, 
%beginning with the Battle of Sulcoit in 967 and 
%culminating in the Battle of Clontarf in 1014.
%{Thus, this period (795-1014) forms part of the Viking Age in Ireland.}
Although its limitations are  well documented,  
the text provides extensive information; it tells of multitudes of characters,  alliances, conflicts, relationships and interactions of all sorts, from a perspective {of} {when} it was written.
Statistical tools to tackle the networks formed by such large casts of characters have recently been developed~\cite{EPL,EPJB,Ossian}. 
Here we apply them in a new investigation to shed quantitative light on the Viking age in Ireland 
%\blue{(for our purposes, approximately ninth to twelfth centuries)} 
as presented in {\CGG}.

%\todo{Network science}
Network science is a broad academic field, related to statistical physics, information visualization, mathematical sociology and other disciplines \cite{AlbertBarabasi,Newman03,Newman_book,Costa2011}.
It enables statistical treatment of certain types of systems comprising large numbers of interdependent elements. 
In character networks, these elements are individual figures (personages), 
represented by nodes (or vertices), and the interactions or relationships between them are represented by edges (or links).
Empirical approaches seek to capture statistics which characterise  such systems \cite{Costa2011}. 
Besides delivering new quantitative insights when applied to old problems, the networks approach inspires new questions and opens new avenues of research.

%\todo{Our aim}
The events associated with the Viking Age in Ireland and Battle of Clontarf are nowadays frequently considered as having entered the public imagination in an overly simplified manner.
That popular picture is  essentially of an ``international'' conflict --- Irish versus Viking --- in which victory for the former ended the latter's ambitions in the country.\footnote{We 
     are aware that {terms related to the word} ``national'' may be viewed as 
		 anachronistic here \cite{Duffyreview2}; we use {them} in the sense of a 
		 large group of people with common characteristics such as language, 
		 traditions, customs and ethnicity \cite{Corrain98} rather than in a 
		 governmental sense \cite{nation}.}
The truth, we are told, is more nuanced and more complex~\cite{Ryan1938,Corrain72}.
Instead of an international conflict, the issue at stake at Clontarf was an internal, domestic, Irish struggle: the determination of Leinster (in the east of Ireland) to remain independent of the dominant dynasties to its north  and south-west~\cite{Ryan1938,Corrain72}.
% Ryan page 49 says ``Fundamentally, then, the issue at Clontarf was the determination of the Leinstermen to maintain their independence against the High''
% O Corrain page 130 says: ``In fact Clontarf was part of the internal struggle for sovereignty and was essentially the revolt of the Leinstermen against the dominance of Brian''
Some such interpretations, wherein the Vikings {are said to have} played a secondary role, tend to downplay the significance of Clontarf \cite{UCD} and have 
been partly ascribed to revisionist fashions \cite{Downham2005,Duffy}.
{{\CGG} has been used to bolster arguments on both sides of the debate.}
Our aim is to determine what its character networks  have to say on the matter.
%Our aim is to determine what the character networks contained in {\CGG} have to say on the matter.

%\todo{Limitations}
{It is important to state from the outset that our analysis is of the content of {\CGG} and its portrayal of the Viking Age in Ireland.
We do not have direct access to the actual social networks of the period and we recognise that the account in 
the {\CG} has been influenced by events and circumstances after 1014 and up to the composition of the text.
We discuss the authenticity and deficiencies of the {\CG} as a source in 
 {Subsection}~\ref{newsec2.2}.
Nevertheless,  the text is important in its own right and, at minimum, tells us how the author sought to represent reality.}

%\todo{Usefulness}
The style of the text of the {\CG} is ``inflated and bombastic'' \cite{Todd}. 
%It is considered by modern scholars to \sout{be a} {contain}  brilliant piece{s} of propaganda whose main purpose was to eulogise Brian Boru and his {\emph{D{\'{a}}l gCais}} dynasty, especially in respect of their claim to the \hl{high kingship} of Ireland \cite{Todd,NMbias,NMannals}. 
It is considered by modern scholars ``as a piece of dynastic political propaganda on 
behalf of the principal lineage of the \DC, the U{\'{\i}} Briain''\footnote{{``U{\'{\i}}'' means ``grandchildren'' or descendants %Duffy page 15
so that the U{\'{\i}} Briain are the descendants of Brian and the  U{\'{\i}}  N{\'{e}}ill are descendants of Niall, etc. ``Ua'' is the singular form.}} \cite{NMbias}.
%page 135
%\mycheck{App. A called.}
(See {{ Appendix~\ref{AppendixA} and Figure~\ref{figureAp1}}} for a brief account of the political structure of Ireland in 1014.)
This is achieved through  extensive and elaborate passages extolling the virtues of Brian and his army while condemning the Vikings as brutal and piratical.
However such qualitative, rhetorical features are largely irrelevant for quantitative character-network analysis. 
Instead, our approach draws only from the most basic  information --- the presence or absence of interactions between characters.
If the text contains networks which are reasonably or approximately reliable in the aggregate, 
they deliver useful information on the society of the {time it presents.}

%\todo{Assortativity}
The entire set of interacting characters in {\CGG} and the relationships between them is represented in {{Figure~\ref{figure1p1}}} of Section~\ref{newsec3}.
The figure represents a network of considerable complexity, similar to those of other epic narratives \cite{EPL,EPJB,Ossian}.
We are  interested in the question whether the {\CG}  networks   are consistent with the traditional depiction of {a} contest which is {clear-cut} international or 
 if they  support the revisionist notion of a power-struggle which is mostly domestic or, indeed, 
 if they deliver something between both pictures.
A simple tally of  edges {(interactions between characters)} will not do as this would not account for different numbers of {Irish} and {Viking} nodes, 
and {a} proper quantitative approach instead necessitates the networks-science concepts of {\emph{assortativity}} and {\emph{disassortativity}}.
The former is the tendency for edges to connect nodes which have similar attributes.
The opposite tendency is disassortativity; whereby links tend to be between nodes of different types. 
The type of attribute we are interested in here is  {{\emph{narrative identity}}}\footnote{The 
      term is motivated by 
			%discussions in {Refs.\cite{Byrne,Corrain98}.
      %In Ref.\cite{Byrne}, Byrne states that ``Like ancient Greece, pre-Norman Ireland 
			%was able to combine political fragmentation with cultural unity'' 
			%% page 8
      %while in 
			a discussion in Ref.\cite{Corrain98} of
			%{\'{O}} Corr{\'{a}}in discusses}
			``the strong sense of identity, achievement, and cultural 
			cohesion that had been created by the Irish learned classes.'' 
			{\'{O}} Corr{\'{a}}in states ``The island was united culturally 
			and linguistically'' and ``Self-consciously, the literati saw the Irish 
			as a people or {\emph{natio}}, to be compared with the Goths, 
			the Franks, or the peoples of 
      classical antiquity. As far as the genealogists were 
			concerned, the Vikings were outsiders, and were called {\emph{Gaill}} 
			`Foreigners' to the end. Irish reaction to the Vikings 
			is to be understood in terms of these cultural 
			traits.'' For further discussions of Hiberno-Scandinavian relations, see {Refs.\cite{NMfoe,Downham2012a}}.} 
 --- categorised as Irish, Viking or other, and taken from the text itself. 
{We} wish to gauge whether nodes linked by different types of edges represent Irish or Viking characters {as presented in the narrative}. 
%\todo{MNM suggests to spell here as ``Ga{\'{\i}}dil''. But this is tricky because ``Gaedhil'' is used in Todd's translation of his title - see 1st sentence in abstract. So delete it?}\sout{(the  \sout{Gaedhil} or the Gaill).}
We  use the {generic} term  {\emph{categorical assortativity}} 
for associated measures which will be used as the primary determinator to distinguish between the alternatives 
listed.
A network with a positive value is said to be {\emph{{categorically assortative}}}.
A negative value  signals disassortativity and 
a value close to zero indicated the absence of any such correlations (neither assortative nor disassortative).

%\todo{Upshot}
We will report that the categorical  assortativity for the conflictual network is moderately negative.
This statistical approach suggests that while the {\CG} account is not as clear cut as either the most traditional or revisionist pictures in the debate depict, it lies on the traditional side.
Thus the networks of {\CGG} give a complex picture of the Viking Age in Ireland comprising predominantly international conflict but with strong degrees of {intranational} hostilities too.
The principal aims of what follows, then, are (i) to present visualisations for the social and conflictual character 
networks, (ii) to {use} the notion of {categorical}  assortativity tailored to estimate where a network of interactions is positioned on the  spectrum from the international to the {intranational} and  (iii) to apply that tool to the networks recorded in {\CGG}.

%%%%%%%%%%%%%%%%%%%%%%%%%%%%%%%%%%%%%%%%%%%%%%%%%%%
\section{{Background}}
\label{sec2}
%\setcounter{equation}{0}
%%%%%%%%%%%%%%%%%%%%%%%%%%%%%%%%%%%%%%%%%%%%%%%%%%%

%\todo{Explain why give background}
Because {\CGG} is a relatively esoteric text (compared with the Greek and Roman classics, for example), in this section, we present a review of existing literature on the topic which it addresses. 
We also discuss the authenticity and deficiencies of {\CGG} as it is used on both sides of the debate.
This review therefore serves to contextualise the text and to motivate a new type of scientific study of it.
%Readers who are only interested in the technical aspects of our work may skip directly to Section XXX.

% % % % % % % % % % % % % % % % % % % % % % % % % % % 
\subsection{{Context:  The war of the Gaedhil with the Gaill}}
\label{newsec2.1}
% % % % % % % % % % % % % % % % % % % % % % % % % % % 

%\todo{3 CGGs + Todd's translation. Different versions are essential for interpolation discussion later.}
{\CGG} comes down to us in three manuscripts. 
The oldest is in the twelfth-century {\emph{Book of Leinster}} 
 which contains  part of the tale.
The second (also incomplete) is the {\emph{Dublin Manuscript}},  dated to the 14th century.
The third and only complete text is the {\emph{Brussels Manuscript.}}
This was { transcribed from an earlier (now lost) manuscript}
%Casey says  ``It is noticeable that although Míchél Ó Cléirigh (one of the Four Masters) 
%had transcribed Cogadh in 1628 and again in 1635, ...
%Downham says ``made by the famous Irish scribe Mícheál Ó Cléirigh. 
%His transcript was made from an earlier copy he made in March 1628 of the now lost ‘Book of Cu Connacht O’Daly’.''
by the famous Franciscan friar M{\'{\i}}che{\'{a}}l {\'{O}} Cl{\'{e}}irigh 
%Denis objected to the word `` save'' but that is how it is presented in Slavin page177: 
%``saving what material on Ireland's past was left in the ancient manuscripts''
who in the 17th century was sent from Louvain in Belgium to Ireland to collect and preserve Ireland's ancient heritage.
The Brussels and Dublin manuscripts are close but not identical.
{M{\'{a}}ire} {\NM} gives a detailed textual history of {\CGG} in Refs.\cite{NMdating,NMannals}.
% In her Breifne Bias paper she basically sais there are 2 recensions.
As a proxy for the orginals, we use the nineteenth-century translation into English by James Henthorn Todd \cite{Todd}.
Todd's edition, which {was} 150 years old in 2017, is accompanied by an extensive introduction and by detailed explanatory footnotes. 
It serves as a source for some scholars wishing to access the narrative today \cite{DuBois}.
Todd considered  {\CGG} as divisible into two parts.
The first recounts the arrival and deeds of the Vikings in  Ireland in a rough chronological fashion.
%{\darkgreen{(Clare Downham discusses the function and chronology of this annalistic section in Ref.\cite{DownhamAnn}.)}}
%The wprd ``rough'' comes form Duffy page 85 - see my notes on him. 
The second part concerns Brian Boru and  his Munster dynasty whose powerbase was on the banks of the river Shannon.
The lives and politics of his family are outlined along with numerous encounters with the Vikings, all leading to the events at Clontarf.

%\todo{Conflict Munster, Meath, Leinster, Dublin}
{{Brian Boru}}  was  king of the  {\DC} in the northern part of the province of Munster (a map of Ireland during the Viking Age is provided in {{Appendix~\ref{AppendixA}}}). 
After various battles at provincial level, Brian and the {\DC} consolidated rule of Munster, defeating their Irish and Norse challengers. Brian then turned his attention to the easterly province of Leinster and the westerly province of Connacht. 
%Stretching from Dalcassians' territory to beyond the central plains, the river Shannon divides Connacht from Leinster and Meath and provided an access route for raiding parties against both east and west.
This brought him into contest with M{\'{a}}el Sechnaill mac Domnaill,  king of Meath and most powerful king in Ireland, but in 997, Brian and M{\'{a}}el Sechnaill 
agreed a truce, whereby the former would rule over the (approximate) southern half of Ireland, while the latter kept the (approximate) northern half. 
By these means, Brian came to control Munster, the area immediately north of {\DC} territory in southern Connacht, 
and Leinster as well as the Hiberno-Norse cities within, 
while M{\'{a}}el Sechnaill held the province of  Meath,  part of Connacht with at least a notional claim of authority over the northern part of Ireland.

%\todo{Munster + Meath against Dublin and Leinster}
In 998, Brian and M{\'{a}}el Sechnaill worked together against the Dublin Norse. 
The Vikings had established a settlement in Dublin in 838 and during the following century they developed a kingdom comprising  large areas surrounding the town and controlling parts of the Irish Sea. 
Viking  Dublin was politically linked at various times to the Isle of Man and the Hebrides, as well as to Viking settlements in Britain and Scandinavia.
Dublin was joined by Leinster under a new king, {{M{\'{a}}el Morda mac Murchada}}, in opposing Brian and M{\'{a}}el Sechnaill. 
Leinster traditionally rejected the rule of both Munster and Meath and the Hiberno-Norse city of Dublin was ruled by M{\'{a}}el Morda's  nephew, {{Sigtrygg Silkbeard}}. 
The two sides met at Glenmama 
%(Cath Ghleann M{\'{a}}ma) in County Kildare 
in late December 999. 
The Irish annals agree that the combined forces of Munster and Meath decisively defeated those of Leinster and Dublin.

%\todo{Ulster and Brian wins Ireland}
The river Shannon presented a barrier to Meath receiving 
support from his ally {{Cathal mac Conchobar mac Taidg}}, 
king of Connacht,  when M{\'{a}}el Sechnaill came under attack by Brian in the year 1000.
By 1002, M{\'{a}}el Sechnaill had submitted to Brian at Athlone \cite{Corrain72}. 
 The next target for Brian was  the northern kingdoms.
It took ten years, a combination of forces and coordinated use of sea and land attacks, and support from the Church in Armagh for the Northern U{\'{\i}} N{\'{e}}ill and regional kings of modern-day Ulster to submit to Brian. 
By 1011, Brian had achieved his aim of bringing all the regional rulers of Ireland under his control.

%\todo{Rebellion}
In 1012, M{\'{a}}el M{\'{o}}rda mac Murchada of Leinster rose in rebellion. 
Allied with {{Flaithbertach Ua N{\'{e}}ill}}, regional king of Ailech in  the north-west, he again attacked Meath. M{\'{a}}el Sechnaill sought Brian's help and the following year Brian and his son led a combined force from Munster and Connacht into Leinster, reaching Dublin in September.  
Out of supplies near the end of the year, they abandoned their siege of the walled city, with an intention to return.

%\todo{Clontarf}
Thus was the background to the famous Battle of Clontarf. 
In 1014, M{\'{a}}el Morda's cousin, Sigtrygg, journeyed to Orkney and the Isle of Man seeking Viking support. 
These Norsemen came under {{Sigurd Hlodvirsson}} (Earl of Orkney, known as Sigurd the Stout) and {{Brodir}}, reputedly  of the Isle of Man.
Brian's forces came from Munster and southern Connacht possibly supported, at least initially, by M{\'{a}}el Sechnaill's Meathmen (the precise role of Meath in the Battle itself is a matter of some contention \cite{Ryan1938,Downham2005,Duffy}). 
%Amongst his Munster contingent he had Hiberno-Norse of Limerick (and perhaps Cork, Waterford and Wexford and maybe Norse mercenaries from Isle of Man?). \mycheck{\red{\tiny{Check explicitly in text!}}} 
%The King of Meath withdrew, however, following a disagreement. \mycheck{\red{\tiny{Check explicitly in text!}}} 
The Battle of Clontarf is believed to have taken place on  Good Friday, 23 April 1014 \cite{Todd}
(see, however, {Refs}.\cite{Duffyreview2,Harrison}). 
%Brian was  also supported by the Scottish forces of {\bf{Domnall earl of Mar}} \cite{Downham}. \mycheck{{\tiny{Check!}}}
%On the opposing side were Sigurd earl of Orkney, who according to Icelandic sagas
%hoped to win kingdom of Ireland \cite{Downham}.
According to the {\CG}, after a day's fighting, the battle ended with the routing of the Viking and Leinster armies.
The account tells us that their retreat was cut off by the high tide.
Many of the nobles died. 
Brodir killed Brian, having found the old man in his tent.
{\NS} informs us that Brodir in turn was killed by {\'{U}}lf Hre{\dh}a 
(possibly Cuduiligh in the {\CG} \cite{Sayers}, meaning Wolf the Quarrelsome), a relative of Brian Boru.
%\red{(In {\CGG}, the character Cuduiligh can translate in a similar way \cite{Sayers}.)}
%\red{(Sayers points out that the name 
%C{\'{u}} Duilig 
%Cuduiligh in {\CGG}
%,  for a member of the Dalcassian personal guard, 
%``corresponds exactly'' with {\'{U}}lf Hre{\dh}a  \cite{Sayers}.)}
%described in Njals Saga  as a brother (or possibly stepson) of Brian Boru but on page 233 of DUFFY, he says Wolf the Quarrelsome was not Brian’s brother but was a family member. He is Úlf hreða. 
Sigurd the Stout of Orkney was also killed, as was the Leinster king M{\'{a}}el Morda mac Murchada. 
Sigtrygg Silkbeard survived and remained king of Dublin, and the king of Meath, M{\'{a}}el Sechnaill mac Domnaill, resumed his claim to  high kingship of Ireland,\footnote{Ireland's most powerful kings were described --- either by themselves, or retrospectively ---
as king of Tara and less commonly, ardr{\'{\i}} (translated as ``high king'').
These concepts were emphasied by the U{\'{\i}} N{\'{e}}ill dynasty who claimed high kingship on the basis of their holding of Tara which long had a special status in Ireland's polity.
%See Duffy page 13, 15
The kingship of Tara rotated
%Duffy page 43.
 between the northern and southern branches of the U{\'{\i}} N{\'{e}}ill 
until {{M{\'{a}}el Sechnaill mac Domnaill}}'s claim to 
the title was interrupted by Brian Boru.
% https://en.wikipedia.org/wiki/List_of_High_Kings_of_Ireland#Historical_High_Kings_of_Ireland
However, assertions of high kingship were just that --- claims rather than unopposed fact.
% duffy page 43: ``partisan assertions rather than statements of undisputed fact.''
The law tracts gave only three grades of king but no ``high king'' or king of Ireland.
% Corrain72 page 28.
For discussion of the nature of kingship and its various grades in Ireland, see, e.g., 
{Refs.}\cite{Byrne,Duffy,Corrain72}.
%Wiki => ``Although the idea of the high-kingship is considered mainly an anachronistic invention, it came into vogue in the 10th century to denote a king who had enforced his power over external territories.[Lalor, Brian, ed. (2003). The Encyclopedia of Ireland. Gill and Macmillan. ISBN 0-7171-3000-2.  page 684]''
}
supported by Flaithbertach Ua N{\'{e}}ill.

% % % % % % % % % % % % % % % % % % % % % % % % % % % 
\subsection{{Authenticity and deficiencies of {\CGG}}}
\label{newsec2.2}
% % % % % % % % % % % % % % % % % % % % % % % % % % % 

%\todo{Advantages of statistical approach}
It is nowadays widely accepted that one of the main aims of {\CGG} was to 
document the achievements of the {\DC} and eulogise Brian Boru
%...
%\red{in support of the pretensions of his descendants, the O Brien kings of Ireland \todo{\black{Check page number in book.}}\cite{Corrain72}}.
%I got this from MNM Dating Considerations paper, page 355.
%...
%In sum, the present scholarly consensus is that {\emph{Cogad}} is a twelfth-century
%composition written at the behest of a direct descendent of Brian B{\'{o}}rama
%who wished 
``\dots to create an illustrious past for his dynasty and to underline thereby later
U{\'{\i}} Brian claims to political power'' \cite{NMdating}.
Although it is a valuable resource for studies of  the Viking Age in Ireland, it is considered a  biased one. 
The question of its  reliability has been the topic of a very long-standing debate~\cite{Todd,Corrain72,NMannals,Ryan1938,Duffy,DownhamAnn}.
Besides some clear interpolation {(described in Subsection~\ref{sec5.2.4})}, much of its bias  appears in the descriptive detail of the narrative. 
Ours, however, is a statistical analysis and, as such, is rather concerned with the totality of the interactions between characters rather than rhetorical levels of detail. 
As with any statistical analysis, what it delivers is a summary which captures aggregate characteristics, largely insensitive to individual elements. 
In this sense, one may hope that it delivers useful statistical information on the Viking Age in Ireland. 
%Even if the outcome is limited at this level, we consider it still interesting to analyse a text of the iconic stature of {\CGG}.

%\todo{The date debate (crucial says Maire)}
Estimates for the date of {\CGG} are various.
Todd stated its author ``was a contemporary and strong partizan of King Brian'' 
\cite{Todd}.
Robin Flower also considered the chronicle  ``almost contemporary'' \cite{Flower1947}.
Albertus Goedheer  gives a date as late as 1160 \cite{Goedheer} but John Ryan argues that {\CGG} ``might have been composed about 1130 or earlier'' \cite{RyanonGoedheer}.
In Ref.\cite{Corrain72}, 
%\blue{\sout{{\'{O}} Corr{\'{a}}in described the {\CG}  as ``concocted ... some two-and-a-half centuries after the date of the events it purports to narrate''.}} \blue{\sout{Elsewhere in}} \blue{\sout{[22]}} 
{Donnchadh} {\'{O}} Corr{\'{a}}in  refers to it as ``written in the twelfth century''.
He also describes the hypothesised text known as {\emph{Brian's saga}} 
 as written about 1100 in response to {\CGG}, a suggestion that implies 
a date before 1100 for the creation of the latter \cite{Corrain72}.
%This is stated on page 91 of Ref.\cite{Corrain72} but on page 46 of the same text he refers to it as a ``twelfth-cetury text'' and on page 78 he says it was ``written in the twelfth century''..
More recent scholarship by {\NM}  gives the  likely composition date of {\CGG} as  between the years 1103 and 1113~\cite{NMdating}.
(She dates the common source for the Dublin/Brussels recension as the 1120s or 1130s 
% See page 361 of the dating paper 
\cite{NMbias,NMdating}.)
%{\darkgreen{Clare Downham considers that {\CGG} ``was composed long after the event'' (i.e. after 1014) \cite{DownhamScot}}}
{Denis} Casey also reviews dating estimates in Ref.~\cite{Casey2013} {and argues that there may have been multiple versions of the {\CG}} (see also Refs.\cite{DownhamAnn,DownhamScot}).
%, sets \red{the date of the recension  in the Brussels manuscript} at about one hundred years after the events described  \cite{Casey2013}.)
Se{\'{a}}n Duffy believes it may be ``based on contemporary annals and, no doubt, local memory''~\cite{Duffy}. 
% Plate 18 of Duffy
He suggests that {\CGG} gives ``a vivid picture of what happened at Clontarf as related perhaps to the writer of the Cogadh by a veteran''
and gives the possibility that it ``was written by someone who may well have lived through these last years of Brian's life''. 
%Duffy Page 214
% Note that in Duffy's talk at https://soundcloud.com/the-royal-irish-academy/stephen-harrison/sets Duffy gives a date of 1080s.
This bringing us back to Todd's original estimate\cite{Todd}.

%\todo{How the debate evolved}
The interpretation of {\CGG} as propagandistic is linked to the question of the date of its composition because ``Heroic stature presupposes nurturing by time'' 
%Page 135
\cite{NMbias}.
Thus its propagandistic nature ``implied that it could no longer be considered contemporary with any of the events it describes'' \cite{NMbias}. 
% (She cites Todd and Flower as two examples of claims that it was contemporary.)
The greater the distance between the events of Clontarf and the setting down of {\CGG}, the more room there is for a distorted view to take hold.
This is the reason why a good estimate date for the composition of the {\CG} is important in the present context. 
Ryan writes:
``In the course of the eleventh century, \dots the view seems to have gained universal acceptance that the Battle of Clontarf was par excellence the great decisive struggle of Irish history. Brian in the retrospect was everywhere acclaimed as a national hero''
%page 47
\cite{Ryan1938}.
The claim is that time distorted reality; 
``The Norse were a substantial section of the opposing force, and in the mellow haze
of popular imagination the battle tended to be transformed into a clear-cut issue, Irish versus Norse, with the former victorious. 
Even in the Northern countries the battle passed rapidly from history into saga'' \cite{Ryan1938}. 
The above estimates for the interval between Clontarf and composition of {\CGG} range between contemporary and {about a} hundred and fifty years. 
Our approach cannot deliver an independent estimate for the date of composition and the above estimates should be kept in mind.
While the above considerations suggest that the {\CG} may distort in favour of an overly international picture of conflict
{(and, indeed, the contemporary name of the tale itself emphasises the Viking-Irish conflict)}, 
on the other hand it should also be kept in mind that, in places, it identifies Leinster as the principal enemies of Brian \cite{Duffy,Duffyreview2}.

%\todo{Interpolation}
In his Introduction to {\CGG}, Todd acknowledges the defects 
of the work and expresses regret that it is ``so full of the feelings of clanship, and of the consequent partisanship of the time, disfigured also by considerable interpolations, and by a bombastic style in the worst taste \dots''.
In chronicle literature, an interpolation of the type mentioned by Todd is a 
later addition not written by the original author.
{We address this issue in Subsection~\ref{sec5.2.4}.}
%** For interpolations see Todd page xvii onwards **

%\todo{The debate about authenticity: CGG is not authentic}
{\'{O}} Corr{\'{a}}in states that the author of {\CGG}~``drew his material from 
the extant annals but he telescoped events, omitted references to other Viking leaders and concocted a super-Viking, Turgesius, whose wholesale raiding and, particularly, whose attack on Armagh was intended to demonstrate the inefficiency of the U{\'{\i}} N{\'{e}}ill as defenders of the church and of the country in contrast of the achievements of the great Brian'' \cite{Corrain72}. (Turgesius is elsewhere referred to as ``exaggerated'' rather than ``concocted'' \cite{Kirwan}.)
{Clare} Downham states that throughout the {\CG}, ``records of alliances between Vikings and
Irish rulers are neglected; a number of victories won by rulers other than U{\'{\i}} Bhriain are omitted.''
Moreover, ``paired names of Vikings rhyme or alliterate and do not transfer easily into Old Norse equivalents 
\dots.  These names look as if they have been invented by the author \dots or drawn from a poetic
source'' \cite{DownhamAnn}.
{{Downham further suggests that since ``historical accuracy, according to the modern definitions, was not the priority'' in {\CGG}, ``the material which is unique to that narrative deserves to be treated with some caution'' \cite{DownhamAnn}.}}

%\todo{The debate about authenticity: CGG is OK}
Duffy, on the other hand 
% Page 124: 
argues that, whatever about the detail of {\CGG} ``and its slightly cavalier approach to chronology'', the gist of the account ``seems sound'' \cite{Duffy}.
Duffy also discusses difficulties in using the annals to check the historicity of {\CGG}. 
%Duffy Page 206
By his reckoning, although some of the names of individuals drafted in from beyond Ireland are indeed suspicious, ``up to half of them appear to be real and their presence at Clontarf is historically credible, if not corroborated by some other source'' \cite{Duffy}.
In Ref.\cite{NMannals}, {\NM} shows that genuine annals underlie {\CGG} and that the compiler of {\CGG} ``remained fairly true to
his exemplar''. 
``Provided, therefore, that we keep the redactor's political purpose firmly in view, we may tentatively
add the annalistic material preserved in {\CGG} to our list of sources for information on the history of Ireland in the Viking
Age'' \cite{NMannals}.

%\todo{Tide - a previous mathematical insight - evidence of reliability}
Todd himself also reports what he considers to be ``curious incidental evidence'' for reliability of at least some of the {\CG} account in that  
it ``was compiled from contemporary materials'' \cite{Todd}. 
``It is stated in the account given of the Battle of Clontarf, that the full tide in Dublin Bay on the day of the battle (23rd April, 1014), coincided with sunrise'' \cite{Todd}.
In a piece of ``mathematical detective-work'' \cite{Duffy} that precedes our own by 150 years, Todd's colleague established that the full tide that morning {occurred} at {5:30~am} and indeed coincided with sunrise. 
For Todd, this ``proves that our author, if not himself an eye-witness, must have derived his information from those who were'' \cite{Todd}.
We have already seen the importance of the time of the evening tide;  
calculated to have been at 5:55~pm, consistent with the account in {\CGG}; it prevented the escape of the Viking forces and considerably aided Brian's victory.  
{(See Ref.\cite{Harrison} for a recent discussion on this topic.)}

This is certainly amongst the most striking evidence in support of the account of {\CGG}. 
Duffy provides multiple other instances where the {\CG} may be reliable \cite{Duffy}.
Certainly  
%Duffy Page 126
bombastic statements that are not backed up by the annals have to be treated warily.
But notwithstanding this, he considers 
%page 123
the narrative as having ``some credibility'', although ``unreliable in its precise detail'' \cite{Duffy}.
(For criticism of Duffy's counter-revisionist views, see e.g., Ref.\cite{Duffyreview2}.)

%\todo{Summary of authenticity issue. Our approach is robust.}
To summarise, there is a vast amount of humanities scholarship concerning {\CGG}. 
Although some dispute its reliability, others consider its version of events mainly credible and largely consistent with other sources and evidence.
As stated by Duffy, 
%Duffy Page 130: The Cogadh, 
``even though it is exaggerated and biased'', {\CGG} can be useful ``if we use it judiciously'' and ``make allowance for its propagandist tendency''.
The composer surely did not think in terms of network science but, in recording  a cast of hundreds connected with well over a thousand links between them, he nevertheless imprinted networks in the narrative.
(We explain how we harvest these data in Subsection~\ref{sec5.1}.)
Thus we may expect that the bulks of the networks contained in {\CGG} might not be too far away from the reality of the networks of the Viking Age in Ireland.
Many of the objections listed above 
are largely irrelevant to our approach as static networks are immune to ``bombastic'' descriptions, ``telescoping'' of events and ``cavalier'' attitudes  to chronology.
We will see that the aggregate approach is even resistant to isolated cases of interpolation.
It is with this perspective that we interrogate the narrative with a networks-science methodology. 
To recap, our primary aim is to determine whether the character networks in {\CGG} are implicative of an ``international contest'' or ``local quarrel'' \cite{IrishTimes}.

% % % % % % % % % % % % % % % % % % % % % % % % % % % 
\subsection{{International contest or local quarrel?}}
\label{newsec2.3}
% % % % % % % % % % % % % % % % % % % % % % % % % % % 

%\todo{The main purpose of our paper - international or intranational?}
Charles O'Connor~\cite{OConnor1766} in the 18th century, with Ryan~\cite{Ryan1938} and {\'{O}} Corr{\'{a}}in~\cite{Corrain72}, in the 20th, are considered early debunkers of the traditional myth of Clontarf \cite{NMdating,Duffy}.
%See http://en.wikipedia.org/wiki/History\_of\_Ireland\_(800–1169)
% says  NMdating on page 355.
O'Connor describes the conflict as a ``civil war'' in which ``the whole  province of Leinster revolted, and called the Normans from all quarters to its assistance''~\cite{OConnor1766}.
%page 264
Ryan's main claim  is that 
%page 7: 
``In the series of events that led to Clontarf it was not \dots the Norse but the Leinstermen, who played the predominant part'' {\cite{Ryan1938}}.
His thesis is that the conflict is not a ``clear-cut'' one between Irish and Viking. 
Firstly, Brian's army was not a national one, but one of Munstermen supported by two small Connacht states. 
Secondly, the opposition ``was not an army of Norse, but an army composed of Leinster and Norse troops, in which the former were certainly the predominant element and may have constituted two-thirds of the whole''~\cite{Ryan1938}.
The battle, then, was not a contest for the sovereignty of Ireland --- it was not a clear-cut issue of Irish versus Norse. Instead, the issue at hand was ``the determination of the Leinstermen to maintain their independence against the High-King''~\cite{Ryan1938}.

It was in the course of the eleventh century, Ryan argues, that the picture of a decisive struggle of Irish history gained ``universal acceptance'' in the popular imagination.
This came about because of the 
%http://en.wikipedia.org/wiki/History_of_Ireland_(800–1169)
parts played by forces from the Isle of Man and the Orkney Islands together with the 
partisan nature of {\CGG}. 
It was only in this retrospect that Brian was acclaimed as a national hero.
{\'{O}} Corr{\'{a}}in's view is similar  \cite{Corrain72}:
``The battle of Clontarf was not a struggle between the Irish and the Norse for the sovereignty of Ireland \dots . [It] was part of the internal struggle for sovereignty and was essentially the revolt of the Leinstermen against the dominance of Brian, a revolt in which their Norse allies played an important but secondary role''.

Duffy points out that this revisionist interpretation is not supported by the other ancient annals.
E.g., the Annals of Inisfallen gives a short but reliable account ``reflective of contemporary reaction to what occurred'' 
%Page 174/175
\cite{Duffy}.
It is stated that ``the Foreigners of Dublin gave battle to Brian'' and Leinstermen are also slain. 
%The brevity of the account is taken by Duffy as a sign of its contemporaneity.
According to Duffy, ``Whereas some modern historians see the Leinstermen as Brian's primary enemy at Clontarf, the annalist was in no doubt that the enemy was the Norse of Dublin. 
%Page 175
In fact he has the same black-and-white picture of the opposing sides that we tend to think of as later legend \dots''.
%, indeed as reflecting a later nationalist, if not xenophobic, outlook.”
``The entry in the Annals of Ulster also echoes the Annals of Inisfallen
% Page 181: 
in emphasising the primacy of the Norse as Brian's adversaries''. 
Duffy states that the Annals of Ulster suggest ``it was fundamentally a contest between the Irish and Norse (although the latter too had Irish allies)''.
%\blue{(Of course, the annalists may also have had propagandistic intentions.)}

Duffy provides multiple items of evidence in support of his view that 
%Page 265
``Brian's principle opponents were the Hiberno-Norse allied to Leinster'' and that the Battle of Clontarf 
``was notable in particular for the great numbers of overseas Norse forces present, and for the huge 
losses they incurred by fighting and drowning''.
``Implicitly, for the Cogadh's author, two centuries of 
Irish opposition to Viking invasion, spearheaded by Brian's dynasty, reached a climax at Clontarf.
That picture was imprinted too, with remarkable correspondences, on the minds of \dots thirteenth-century Icelandic writers. Those who did battle with Brian came from the Norse world seeking a kingdom for themselves in Ireland''.

Thus, the debate about Clontarf has spanned the centuries and frames our present investigation.
Here we broaden the question to how conflictual and social relationships are presented in {\CGG}.

%%%%%%%%%%%%%%%%%%%%%%%%%%%%%%%%%%%%%%%%%%%%%%%%%%%
\section{{Methods: The {\CG} narrative network}}
\label{newsec3}
%\setcounter{equation}{0}
%%%%%%%%%%%%%%%%%%%%%%%%%%%%%%%%%%%%%%%%%%%%%%%%%%%

In this section, we explain the methods by which the data were harvested and our focus on network topology.
We also present a visualisation of the {\CG} narrative network and discuss how interpolation has negligible effect on our network statistics. 
To keep the main text manageable, we defer details concerning various assortativity measures to Appendix~\ref{AppBassort} and the roles played by the most important characters to {{Appendix~\ref{AppDnames}}} along with an analysis of network robustness.

% % % % % % % % % % % % % % % % % % % % % % % % % % % 
\subsection{{Constructing the {{\CG}} network}}
\label{sec5.1}
% % % % % % % % % % % % % % % % % % % % % % % % % % % 

%\todo{Intro and staving off attack by humanities experts}
As with previous studies \cite{EPL,EPJB,Ossian,PMCthesis,ACS}, we consider {\CGG}
as playing out on a complex  network comprising $N$ nodes and $M$ edges.
The edges link the nodes through  relationships or interactions.
We distinguish between three categories --- Irish, Viking and other --- identifying to which group each node belongs from the text itself. 
%In the text, the first two groups are presented starkly as in binary opposition to one another and since our analysis is of the text we take its data at face value.
We obviously cannot directly access the reality behind the text to determine any gradation between the groups. 
For example, we cannot know how Sigtrygg Silkbeard, who had a Viking father and an Irish mother, might have self-identified in reality; we can only take our lead from the {\CG} itself and since the Hiberno-Norse of Dublin are presented there as Vikings, they are placed that category. 
Nodes classified as ``other'' are those that are not readily assigned to either camp.

%...................................................................................
\begin{figure}[t]
\begin{center}
\includegraphics[width=0.7\columnwidth,angle=0]{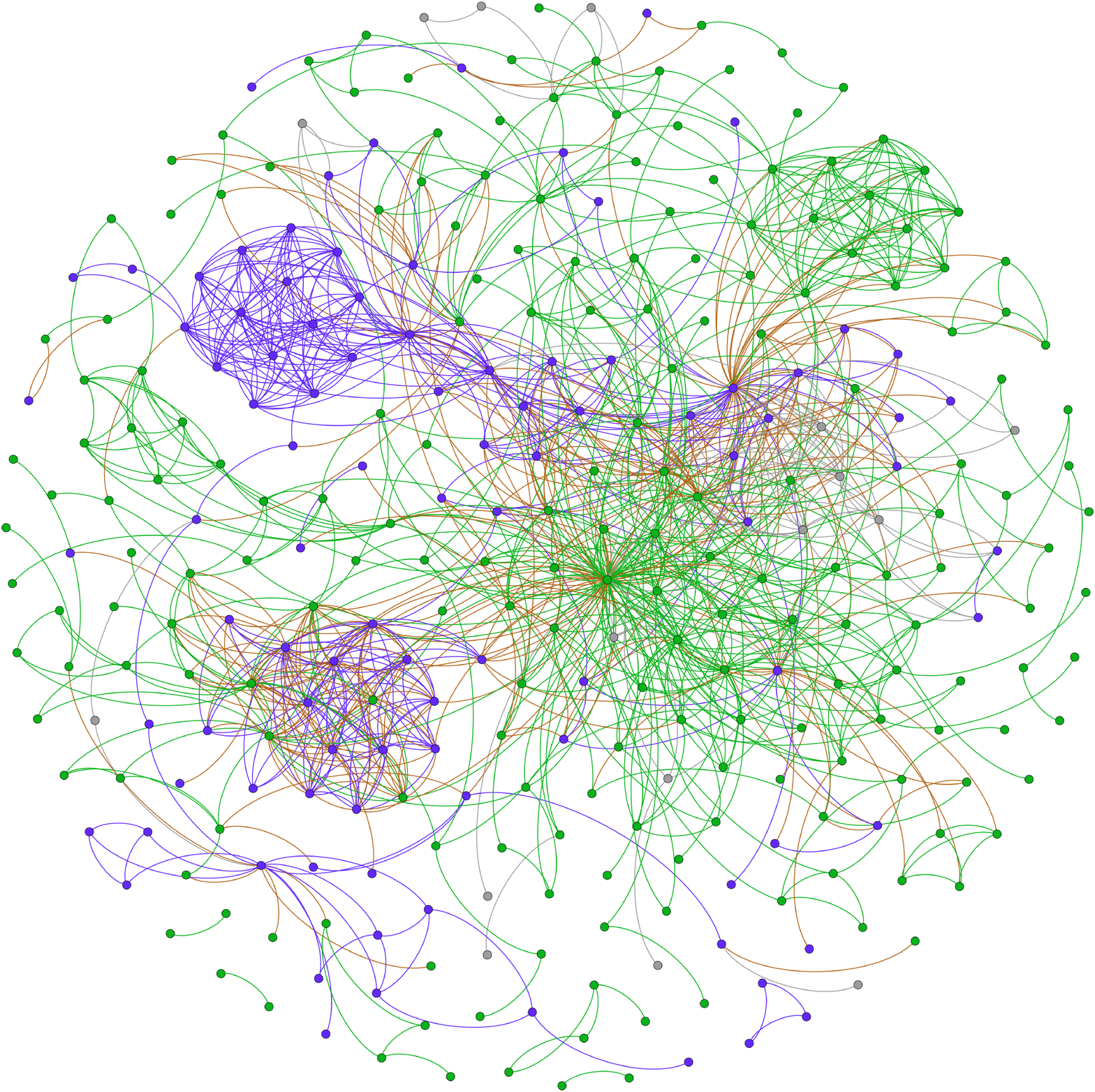}
\caption{The entire {\CG}  network of interacting characters. 
%In this unsigned, full-cast  network (i.e., including  all interacting characters and both positive and negative edges), c
Characters identified as Irish are represented by green nodes and those identified as Vikings are in blue. Other characters are in grey. 
Edges between pairs of Irish nodes are also coloured green while those between Viking pairs are blue. 
Edges linking Irish to Viking nodes are brown and the remaining edges are grey. 
}
\label{figure1p1}
\end{center}
\vspace{0.3cm}
\end{figure}
%...................................................................................

%\todo{Have to manually gather data}
 {Our} approach to constructing the networks follows the methodology of Refs.\cite{EPL,EPJB,Ossian} in that nodes and links are identified by carefully and manually reading the texts with multiple passes through all of the material by multiple readers. 
In our experience, such an approach is required to minimise errors and omissions as well as to reduce levels of subjectivity.
{\CGG} is a very dense text and meticulous care is required to interpret extremely subtle tracts containing large amounts of explicit and implicit information.
It is currently beyond  technological capabilities to extract such information automatically owing to the inherent complexity of such texts (see, e.g., Ref.\cite{Marcello}).
Establishing the technology for such an approach is another active area of research.

%\todo{Visualisation}
{{Figure~\ref{figure1p1}}} contains a network visualisation of the full set of interactions recorded in the {\CG}.
Green nodes represent Irish characters and green edges represent interactions between them.
The counterpart set of Viking nodes and their interlinks are in blue.
Brown edges represent interactions between Irish and Viking nodes.
Any remaining nodes and edges are in grey.

%\todo{Positive vs negative}
We distinguish between two types of edge: positive and negative.
Positive edges are established when any two characters are related, 
communicate directly with each another, or speak about one another, 
or are present together when it is clear that they know each other. 
So positive edges ordinarily represent familial or social relationships.
Negative links, on the other hand,  are formed when two characters meet 
in physical conflict or when animosity is explicitly declared by one 
character against another and it is clear they know each other (such as declarations of war).
So negative edges typically represent actual or intended physical hostility.
It is possible that two characters are linked by both positive and negative 
edges as relationships between characters may change over time.

%\todo{METHODOLOGY: Why static method is OK. Taken from page 15.}
Ours is a static analysis, capturing the temporal totality of the {\CG} narrative.
``Making the past just as visible as the present'', as Moretti puts it \cite{Moretti},
is a  benefit of this networks approach and one which has been used elsewhere~\cite{EPL,EPJB}.
Nonetheless, it should be noted that the study of dynamical properties of networks constitutes an active, broad and developing area of research and such an approach would be of interest in the future \cite{ACS}.
%\todo{Why non-weighted network is OK. From page 15}
We focus primarily on the {\emph{topology}} of the networks underlying  {\CGG},  considering undirected, unweighted networks.
This means that (i) the features which connect the various nodes are not oriented and (ii) the statistics we report upon do not take into account varying levels of intensity of interactions between nodes. 
To account for (i), one would have to introduce a level of detail which is finer that just positivity or negativity.
However, what one gains in refining details, one loses in statistical power. 
To account for (ii), one may place higher weight on more intense interactions,
but, besides using the number of interactions between characters in the narrative, there is no established standard mode of weighting edges in character networks. 
Moreover, we are primarily interested in the presence or absence of conflict, not on the details of varying intensity of such hostility.
Therefore we defer consideration of directed, weighted and temporal networks for future studies and restrict the current study to network topology and related matters.

% % % % % % % % % % % % % % % % % % % % % % % % % % % 
\subsection{Network methodology: basic statistics}
\label{newsubsec3.2}
% % % % % % % % % % % % % % % % % % % % % % % % % % % 

%\todo{Assortativity}
We identified  {$N=315$} individual interacting characters in  Todd's translation  of {\CGG}.\footnote{Actually, 
         we identified {326} individual characters in total.
				 Of these,  {11} are isolated in the sense that they do not interact in the narrative.
				 We consider these  as not forming part of the {\CG} network and they are omitted from our analysis.
				 The characters were identified in the main part of Todd's text. 
				 Todd's paratexts (introduction, footnotes, appendices and index)
				 were used to aid the identification of characters and links between them but 				 
				 individuals mentioned only in the paratexts  do not form part of the {\CG} network. 
				 A small number of characters appear in the main text but are omitted in  Todd's index.
				 We also identified {34} groups of unnamed characters. If considered as nodes, 
				 they bring an additional {187} edges.
				 However because these are neither individuals or named, we omit them from our presentation too.
				 Besides, and for completeness, we also analysed the networks with these nodes included and they 
				 deliver only very small changes to the statistics 				 presented here. }
				These nodes are interconnected by  {$M=1190$} edges
and we refer to the corresponding assemblage as the {\emph{entire}} network.
We can also consider the positive and negative sub-networks,  
formed only of positive or negative edges, respectively.
Examination of these allows us to gain more insight into the social and conflictual statistics contained in the narrative.
Indeed, it is long known from sociology that societies exhibit {\emph{homophily}}, the tendency of individuals to associate with others who are similar to themselves \cite{homophily,Newman2002,Newman2003}.
In the field of social network analysis, this is known as assortativity.
In previous studies of epic literature \cite{EPL,EPJB,PMCthesis,Ossian}, we studied {\emph{degree assortativity}}, the tendency (or otherwise) of nodes to attach to other nodes with similar numbers of links. 
We found  some positive sub-networks exhibit degree assortativity, or are uncorrelated, while the opposite feature --- degree disassortativity --- is characteristic of negative sub-networks. 
This means that positive social networks  give a ``cleaner''   picture (relative to full networks) of the non-conflictual societies underlying such narratives, making it valuable to study them in isolation  \cite{NewmanPark}.
A new feature of the current study is our additional focus on the negative sub-network to statistically measure levels of hostility.

\begin{table}[t]
\caption{Full-cast networks comprise  Irish, Viking and other nodes together with interactions between them.
Unsigned networks comprise   positive and negative edges as well as the nodes they connect. 
Thus, for example, the positive, full-cast network comprises all nodes but only positive links.
The unsigned, Irish network comprises only Irish nodes but both positive and negative links between them. 
The {\emph{entire}} network comprises all interacting nodes and all links.
}
\begin{center}
\begin{tabular}{|c|l|c|c|cc}  
\cline{1-4}
  \multicolumn{1}{|c}{}  &            &    \multicolumn{2}{c|}{Edges}         &    & \\
\cline{3-4}
  \multicolumn{1}{|c}{}  &            & Positive  & Negative  &  &  \\
\cline{1-4}
\multirow{3}{*}{\rotatebox{90}{{Nodes~}}}   & Irish      &           &           &  \multirow{3}{*}{\rotatebox{90}{{$\underbrace{~~~~~~~~~~~}$}}} & \multirow{3}{*}{\rotatebox{90}{{Full cast}}} \\
\cline{2-4}
   & Viking     &           &           &    & \\
\cline{2-4}
   & Other      &           &           &    & \\
\cline{1-4}
 \multicolumn{1}{c}{}  & \multicolumn{1}{c}{} & \multicolumn{2}{c}{$\underbrace{\quad\quad\quad\quad\quad\quad\quad\quad}$}  & &\\
 \multicolumn{1}{c}{}  & \multicolumn{1}{c}{} & \multicolumn{2}{c}{Unsigned}       & &\\
\end{tabular}
\end{center}
\label{table5p1}
\end{table}

%\todo{Terms}
We use the term {\emph{unsigned}} to refer to networks containing both positive and negative edges. 
Networks comprising only positive (or only negative) edges are then themselves termed {\emph{positive}} (or {\emph{negative}}, respectively).
We use the term {\emph{full-cast}} to refer to networks containing the full cast of characters, Irish, Viking and others.
Networks containing only Irish (or only Viking) characters are themselves referred to as {\emph{Irish}} (or {\emph{Viking}}, respectively).
This terminology is summarised in {{Table~\ref{table5p1}.}}
Statistics for the entire network and various sub-networks are collected in {Table~\ref{table5p2}.}

\begin{table}[t]
\caption{Statistics for the entire  network and its various sub-networks. 
The first and second columns indicate whether the sub-network is unsigned, positive or negative with full cast of characters (Irish, Viking and other)
or only the Irish or Vikings are taken into account.
Here, $N$ represents  the number of nodes;
$M$ is the number of edges;  
$\langle k \rangle$ is the mean degree and $k_{\rm{max}}$ its maximum.
The proportion of  triads  that contain an odd number of positive links is represented by $\Delta$
and the 
degree assortativity is denoted by $r$.}
%\vspace{2ex}
\begin{center}
%\resizebox{\textwidth}{!}{%
\begin{tabular}{|l|l|r|r|r|r|r|r|}  \hline
        &           & $N$~ &$M$~   &${\langle{k}\rangle}~$ 
				                               & $k_{\rm{max}}$   &$\Delta$~~& $r$~~~~           
\\  \hline  
        &           &     &      &     &           &             &   \\
\multirow{3}{*}{\rotatebox{90}{{Unsigned}}}
~       & Full cast & 315 & 1190 & 7.6 & 105       &  0.93       &  -0.09(2)  \\
        &Irish      & 193 & 530  & 5.5 &  63       &  0.93       &  -0.08(3) \\
        & Vikings   & 91  &  313 & 6.9 &  26       &  1.00       &  0.31(7)  \\ 
        &           &     &      &     &           &           &   \\
\hline 

        &           &$N^+$&$M^+$ &${\langle{k}\rangle^+}$ 
				                               &$k^+_{\rm{max}}$& & $r^+$~~~~    
\\  \hline  				
        &           &     &      &     &                               &           &   \\
				\multirow{3}{*}{\rotatebox{90}{{Positive}}}
		    & Full cast & 287 &  957 & 6.7 &  53             &           &  0.00(4)  \\ 
        & Irish     & 186 &  475 & 5.1 &  47             &            &  -0.02(4)  \\
        & Vikings   &  88 &  301 & 6.8 &  26             &            &   0.34(7)  \\ 
        &           &     &      &     &                 &         &   \\
\hline 
        &           &$N^-$&$M^-$ &${\langle{k}\rangle^-}$ 
				                               &$k^-_{\rm{max}}$&   & $r^-$~~~~    
\\  \hline  				
        &           &     &      &     &           &              &       \\
\multirow{3}{*}{\rotatebox{90}{{Negative}}}
		    & Full cast & 180 &  264 & 2.9 &  63          &          &  -0.25(3)  \\ 
        & Irish     &  62 &   72 & 2.3 & 25              &          &  -0.26(6)  \\
        & Vikings   &  18 &   16 & 1.8 & 4              &         & -0.08(18)  \\ 
        &           &     &      &     &                       &            &   \\
\hline

\end{tabular}
%}
\end{center}
\label{table5p2}
\end{table}

%\todo{Stats}
The average number of edges per node for the entire network is  {$\langle{k}\rangle = 2M/N \approx 7.6$}. 
The actual number of edges associated with the $i$th node is denoted by $k_i$. 
This is a number which varies between $1$ for the least connected characters (nodes with $k_i=0$ have no links and are not attached to the network at all) and $k_{\rm{max}}$ for the most connected (in a sense, the most important) character.
For the entire network, the most connected character is Brian himself who, with  {$k_{\rm{max}} =  105$} edges, is linked to  {33\%}  of the other characters in the narrative.
Besides Brian's degree, we are also interested in the connectedness of other characters and we  rank the first few characters according to their individual degrees, and according to other measures of importance, in Appendix~\ref{AppDnames}.

%\todo{+/- stats}
{\CGG} has {$N^+=287$} interacting characters in its positive sub-network, 
interconnected by {$M^+=957$} edges, corresponding to a mean degree of  {$\langle{k}\rangle^+ \approx 6.7$}.\footnote{Again 
       we have omitted isolated nodes from the 
			 positive and negative sub-networks. 
       % We do this because the negative sub-networkcan have a large number od zero-nodes and may end up with average degree less than 1, which is not sensible for comparisons to other networks.
}
Here and henceforth, we use the superscripts ``$+$'' and ``$-$'' to identify statistics associated with the positive and negative networks, respectively. 
(We omit such a superscript from statistics for the unsigned networks.
These are distinguished from generic symbols by context.)
The counterpart  figures for the negative network are  {$N^-=180$},  {$M^-=264$} and {$\langle{k}\rangle^- \approx 2.9$}, respectively.
(The total number of positive and negative links  {$M^+ + M^-=957+264 = 1221$} exceeds the number  {$M=1190$} which we previously identified for the entire network because some relationships involve both positive and negative aspects.)
As for the entire network, Brian has the highest degrees in both positive and negative subgraphs, with the former measured at {$k^+_{\rm{max}} = 53$} and the latter at {$k^-_{\rm{max}} = 63$}.

%\todo{$\Delta$}
The adage that ``the enemy of an enemy is a friend'' is related to the notion of 
{\emph{structural balance}} in network science~\cite{Heider,Cartwright,Antal}. 
The maxim suggests that {\emph{triads}} (sets of three mutually connected nodes) with one positive and two negative edges are commonplace.
More generally, triads with odd numbers of positive edges are considered structurally balanced.
One way to quantify the extent to which it holds in a character network is through the statistic $\Delta$, defined  as the percentage of triads that contain an odd number of positive links. 
A large value of $\Delta$ means that  hostility between two characters is suppressed if they have a common foe. 
Clearly $\Delta$ is only meaningful for the unsigned network; on the positive sub-network it is 1 by definition, while in the negative sub-network it is necessarily zero.
We find that the entire network underlying {\CGG} 
(which has  {3041} triads) is indeed structurally balanced with  {{$\Delta \approx 93\%$.}}

%\todo{$r$}
As mentioned above, assortativity (disassortativity) is the tendency for the nodes of a  network  to attach to other nodes that are similar (different) in some way. 
Network theorists frequently measure {\emph{degree assortativity}} --- the extent to which nodes of similar degree tend to link up.
As with other character networks, we find that the negative full-cast network is disassortative by degree {$r = -0.25(3)$}].\footnote{The error here 
      is estimated using the method described in Refs.\cite{Newman2002,Newman2003}. 
			Error estimates for other network statistics are small 
			(see discussion in the final paragraph of Subsection~\ref{sec5.2.4})
		  and we refrain from reporting them here.
      We only display assortativity errors because they provide useful 
			information when comparing systems which are, or nearly are,
			uncorrelated ($r$ close to zero).}
This means that high-degree characters are hubs and their negative links preferentially attach to  low-degree ones.
This appears to be a generic feature of heroic tales in particular, where the hero or heroes encounter multitudes of lesser characters and defeat them in battle.
The positive full-cast network, on the other hand, is uncorrelated within errors [{$r=-0.00(4)$}, meaning it is neither assortative nor disassortative]. 
These features are typical of social networks and of character networks with positive interactions \cite{NewmanPark,EPL}.

%\todo{Irish and Viking networks}
Beside the networks comprising the full cast of characters, we can also consider the networks containing only Irish or only Viking nodes and these are also listed in  {{Table~\ref{table5p2}.}}\footnote{As usual, isolated 
          (degree-zero) nodes are  removed.
					E.g., there are 202 Irish nodes in total (see Table~\ref{table5p4}), 
					but 9 of these are disconnected from other Irish nodes, so they are omitted 
					from the unsigned Irish network in Table~\ref{table5p2}.
					Besides the value of $N$, reinstating them does not alter the statistics listed within the precision of Table~\ref{table5p2}.
					}
We observe the following average properties of the various networks.
In the  Irish, and Viking networks (as in the full-cast cases), the mean degrees are {maximal} for the unsigned networks and {minimal} for the negative sub-networks.
The unsigned Viking network is more structurally balanced than its Irish counterpart. 
Structural balance  for the Irish network, which has  {830} triads, is  {93\%}
whereas the  {881} Viking triads  {all} contain odd numbers of positive links. 
% All of this suggests that the Irish networks are more similar to the fill-cast networks than are the Viking ones.

%  %  %  %  %  %  %  %  %  %  %  %  %  %  %  %  %  %  
\subsection{{Effect of interpolation on network statistics}}
\label{sec5.2.4}
%  %  %  %  %  %  %  %  %  %  %  %  %  %  %  %  %  %   

%\todo{What is interpolation}
In his Introduction to {\CGG}, Todd acknowledges the defects 
of the work and expresses regret that it is ``so full of the feelings of clanship, and of the consequent partisanship of the time, disfigured also by considerable interpolations, and by a bombastic style in the worst taste \dots''.
In chronicle literature, an interpolation of the type mentioned by Todd is a 
later addition not written by the original author.
As scribes copied ancient material by hand, extraneous material frequently came to be inserted for a variety of reasons \cite{McCarthy2008}.
These may have been for bona fide intentions, perhaps as explanations; 
for subjective purposes; or they may simply have crept in through errors and inaccuracies arising from manual copying or, indeed, as attempts ``to enhance the appeal of the narrative'' \cite{NMbias}.
%I got the last one from page 141 of the Brefnie Bias paper 
One way to detect such interpolation is through comparing different manuscripts.

%\todo{Ua Ruairc}
Perhaps the most famous interpolation in the narrative is a passage 
which occurs in the Dublin version  describing the actions 
of Fergal Ua Ruairc  of Br{\'{e}}ifne and associate chieftains~\cite{Ryan1938,NMbias}.
(For the location of Br{\'{e}}ifne, see Figure~\ref{figureAp1}.)
% Domhnall, son of Ragallach (Reilly) and Gilla-nanaemh, son of Domhnall, and grandson of Fergal,
The Brussels manuscript,  by contrast, ``omits everything connected with Fergal and 
his presence in the battle'' \cite{Todd}.
As stated by Todd, ``the whole story bears internal evidence of fabrication, for Fergal O'Ruairc was slain A.D. 966 
\dots,
%[964, Four M.], 
and our author had already set him down amongst Brian's enemies''. 
Ryan \cite{Ryan1938}, Duffy \cite{Duffy} and others also identify Ua Ruairc as an interpolation and N{\'{\i}} Mhaonaigh gives a detailed account of Br{\'{e}}ifne bias in {\CGG} \cite{NMbias}. 
She states ``one of the main aims of the interpolator was to portray Fergal Ua Ruairc and his followers in as favourable a light as possible, sometimes regardless of the effect this had on his text''.
The point is that {a} pro-Ua Ruairc reviser of the narrative may have deemed it politically expedient to alter the record of relations between  the U{\'{\i}} Ruairc and the {\DC} by demonstrating assistance given by the former to Brian at Clontarf. 
N{\'{\i}} Mhaonaigh estimates the period when the U{\'{\i}} Ruairc were  likely to have gained maximum advantage from such an association to have been the mid- to late 1140s, over a hundred years after Clontarf \cite{NMbias}.

%\todo{Ua Ruairc is structurlally imbalanced}
{We are interested in what insight the networks methodology can give on such matters.}
We have already seen that  {{93\%}} of the  {3041} triads in the unsigned network are structurally balanced
as are  {{93\%}} of the {830} triads in the Irish network.
The triad formed by Ua Ruairc's enmity to M{\'{a}}el Sechnaill, the latter's alliance with Brian, and the interpolated support of Ua Ruairc for Brian is one of two positive edges and one negative one, which is structurally {\emph{imbalanced}}. 
Since the vast majority of triads in {\CGG} are balanced, this makes the Ua Ruairc episode stand out as relatively unusual. 
We removed Ua Ruairc and his three associates 
(Gilla-na-Naomh,  Mac an Trin and
Domhnall mac Raghallach \cite{Todd}) from the networks  to test the effects on the  statistics. 
% Domhnall, son of Ragallach (Reilly) and Gilla-nanaemh, son of Domhnall, and grandson of Fergal,
Besides reducing  the number of edges (e.g., $M$ reduces from {1190} {to 1146} in the entire network), the effects of this removal are minimal.
For example, the degree assortativies are unchanged within error estimates for the unsigned, positive and negative networks.

%\todo{Randomly remove nodes/edges}
The possibility of interpolation applies not only to Ua Ruairc and allies.
Ryan claims that ``Many of the names mentioned are names only, for nothing is known of the persons who bear them. 
Some of the levies in important positions were certainly absent. In a word, no effort is  made to distinguish between the genuine and the spurious, to criticise suspect sources, and to reconcile contradictions''~\cite{Ryan1938}.
Given the minor effect of the most famous and easily identified, Ua Ruairc, interpolation,
% the U{\'{\i}} Ruairc material is  how it is expressed in the Breifne Bias paper.
we do not attempt to remove other interpolations from our analysis.
Besides, any attempt to do so would be incomplete because we cannot be certain that all interpolations have been identified.
Indeed, as we have repeatedly emphasised, ours is a network study of {\CGG} as represented by Todd
%Duffy Page 126: something about the expectations of the time \cite{Duffy}.
 in Ref.\cite{Todd} and therefore we present  it in its entirety.
However, we attempt to simulate the effects of interpolation by randomly removing up to $15\%$ of nodes or edges. 
The process is repeated 1000 times and the averages deliver no appreciable difference to the statistics 
given in {Table~\ref{table5p2}}, indicative of their robustness (see {{Appendix~\ref{AppDnames}}} for a network-robustness analysis).
For example,  removal of 15\% of the vertices alters the assortativity  from $r = -0.09$ to $r = -0.08$ (imperceptible change within errors).
Removal of 15\% of the edges   leaves $r$ unchanged within this level of precision. 
A more systematic and targetted quantitative study of the effects of interpolation would be interesting for future study.

%%%%%%%%%%%%%%%%%%%%%%%%%%%%%%%%%%%%%%%%%%%%%%%%%%%
\section{Results: The relationships between Irish and Vikings as recorded in the {\CG} networks}
\label{newsec4}
%\setcounter{equation}{0}
%%%%%%%%%%%%%%%%%%%%%%%%%%%%%%%%%%%%%%%%%%%%%%%%%%%

%\todo{Aim}
The traditional ``memory'' of the events leading up to the Battle of Clontarf is of an international conflict between two distinct sides: Irish vs Viking \cite{Ryan1938}.
This is dismissed by revisionist historians who argue that the conflict is primarily Irish-on-Irish \cite{OConnor1766,Ryan1938,Corrain72}.
The traditional viewpoint of a clear-cut contest  might be expected to lead to a network in which the bulk of negative (conflictual) edges correspond to Irish-Viking interactions representing the primacy of hostility being between the two groups. 
We might expect a network supporting the revisionist stance to be somewhat different: the negative edges would mainly link Irish nodal pairs.  
We also have to monitor Viking-on-Viking conflict 
as there were different Viking factions in Ireland during this period~\cite{Todd,Smyth}.

%%%%%%%%%%%%%%%%%%%%%%%%%%%%%%%%%%%%%%%%%%%%%%%%%%%%%%%%%%%%%%%%%%%%%%%%%%%%%
\begin{table}[!b]
\caption{Identity profiles of the cast and their interactions in {\CGG}. 
The second, third and fourth rows give the numbers (and percentages) of nodes which are identified as Irish, Viking and other (not identified as Irish or Viking) in the entire, unsigned network as well as in the positive and negative sub-networks. 
The fifth row gives the total number of nodes in each network (these values are $N$, $N^+$ and $N^-$ for the full-cast networks, respectively).
The sixth and seventh rows give the numbers (proportions) of edges which connect pairs of like nodes.
The eighth row gives the numbers (proportions) of edges which connect Irish and Viking nodes. 
The last row gives the total numbers of edges in each case as ($M$, $M^+$ and $M^-$ for the full-cast networks).
The remaining edges involve other (not assigned as Irish or Viking) nodes.}
\begin{center}
\begin{tabular}{|l|r |r | r |}
\hline 
                      & \multicolumn{1}{c|}{Entire}    & \multicolumn{1}{c|}{Positive}  & \multicolumn{1}{c|}{Negative}  \\ 
                      & \multicolumn{1}{c|}{network}   & \multicolumn{1}{c|}{network}   & \multicolumn{1}{c|}{network} \\ 
\hline 
Irish nodes           & 202 (64 \%)       &  187 (65 \%)      &    110 (61 \%) \\ 
\hline 
Viking nodes          &  97 (31 \%)       &  88 (31 \%)       &    61 (34 \%) \\ 
\hline 
Other      nodes      &  16 (5 \%)        & 12 (4 \%)         &    9 (5 \%) \\
\hline
{\bf{Total \# nodes}} & 315 (100 \%)   		& 287 (100 \%)		  &  180 (100 \%) \\
\hline \hline
Irish-Irish   edges   & 530 (45 \%)       & 475 (50 \%)       &     72 (27 \%) \\ 
\hline 
Viking-Viking edges   & 313 (26 \%)       & 301 (31 \%)       &     16 (6 \%) \\ 
\hline 
Irish-Viking  edges   & 272 (23 \%)       & 119 (12 \%)       &     163 (62 \%)  \\ 
\hline
{\bf{Total \# edges}} & 1190 (100 \%)       & 957 (100 \%)        &     264 (100 \%) \\
\hline
\end{tabular} 
\label{table5p4}
\end{center}
\end{table}

%\todo{Figs and tables}
In Table~\ref{table5p4}, we record the proportions  of Irish, Viking and other nodes in the unsigned  networks and in its positive and negative sub-networks.\footnote{Some of the entries in the second and third rows of Table~\ref{table5p4} differ from entries in the third column of Table~\ref{table5p2} because isolated nodes are not removed from sub-networks in Table~\ref{table5p4}.
This is because Table~\ref{table5p4} concerns identity profiles of unsigned, positive and negative networks, in distinction to the  Irish and Viking sub-networks of Table~\ref{table5p2}. Numbers of edges match across both tables, however, because, by definition, these do not involve isolated nodes.}
At  {61\% -- 65 \%}, the proportions of Irish nodes in each of the three graphs are approximately constant.
The proportion of Viking nodes is also relatively stable  between  {31\% and 34\%.}
%\todo{Exploratory results}
In the same table, we list the proportions of interactions which link Irish to Irish nodes; Viking to Viking; and Irish-Viking pairs.
{Fifty percent} of edges in the positive network  link pairs of Irish nodes; {31\%} connect  pairs of Viking nodes;
 and   {12\%} of positive interactions connect mixed Irish-Viking pairs.
 {Twenty-seven percent} of links in  the negative network   connect Irish to Irish nodes; 
 {6\%} connect pairs of Viking nodes;
and over  {62\%} of negative interactions connect mixed Irish-Viking pairs.
In other words, the positive (social) network   is dominated by interactions between characters of the same narrative identities
({{intranational}} interactions) 
and the  negative (conflictual) network   is dominated by Irish-Viking  (international) interactions.
This suggest that the  largest proportion of {\CG} conflict is international, but there are significant levels of {intranational} hostilities too (especially Irish versus Irish).
Actually, from {{Table~\ref{table5p4},}} we see that the number of international edges in the negative network is over twice the number of Irish-Irish negative edges, which, in turn is over four times the number of Viking-Viking negative edges.

%\todo{How to do it properly}
However, to properly evaluate the levels of mixing, negative or positive, between Irish and Viking, one has also to account for the fact that they do not have the same numbers of nodes in the networks
(there are twice as many Irish nodes as Viking). 
To do this, we introduce the {\emph{categorical assortativity}} of the various networks, represented generically by $\rho$.
Its precise definition is given  in {{Appendix~\ref{AppBassort}.}}
It is a measure which ranges between ${\rho}_{\rm{min}}$ and 1 where 
$ {\rho}_{\rm{min}} $ is a non-trivial, negative value, which itself lies between $-1$ and $0$ if there are more than two categories under consideration~\cite{Newman2002,Newman2003}. 
Thus, although the maximum value of  $\rho$ is one, its minimum value can be network-dependent.
The reason for this is that, when there are more than two categories, disassortativity connects dissimilar nodes, just as randomness does. 
Assortativity, however, connects like nodes and is therefore quite different to randomness. 
We have to be mindful of this asymmetry when interpreting the categorical assortativity for the negative networks with three categories of node (Irish, Viking and unassigned).
The only instance in which $ {\rho}_{\rm{min}} =-1$ is when there are  two categories.

%\todo{Positive nw results - $\rho^+$. Trim?}
The value $\rho=1$ indicates 100\% categorical assortativity.
If this were the case for our positive network, for example, it would mean that the only positive interactions are {\emph{within}} rather than between categories
(friendly interactions would be {intranational}). 
The value {$\rho={\rho}_{\rm{min}} < 0$} implies that the network is fully categorically disassortative.
If this were the case for our positive network it would mean that the only positive interactions are {\emph{between}} rather than within categories
(positive interactions would be international).
A value $\rho=0$ would indicate that the categorical assortativity is the same as would be expected for random mixing between nodes, oblivious of their Irish or Viking character.
We find that {$\rho^+ = 0.65(3)$} for the full-cast positive network.
If we restrict our attention to Irish and Viking nodes only by removing other nodes, this rises to {$\rho^+ = 0.72(3)$.}
These statistics are recorded in 
{{Table~\ref{table5p5}}} and support the picture that most (but not all) positive interactions are {intranational}.

%\todo{Negative nw results - $\rho^i$. Trim?}
We now focus our attention on the negative networks as these connect with the debate in the humanities discussed in Section~2.  
A ``clear-cut'' version of the ``international-conflict'' picture  would  be characterised by the value $\rho^- \approx {\rho}_{\rm{min}}^- $ (where ${\rho}_{\rm{min}}^-$ is the minimum possible value of $\rho^-$, and  is $-1$ when unassigned nodes are excluded). 
Such a value would reflect a purely Irish-versus-Viking conflict.
At the opposite end of the spectrum would be a world in which all conflict is {intranational}. 
In this case one would expect $\rho^- \approx 1$.
The revisionist picture of a primarily (but not exclusively) {intranational} conflict may be expected to correspond to a positive value of $\rho^-$.
Between the two extremes, we might imagine a more even distribution of negative edges, whereby conflict between nodes is ``blind'' to their identities. 
A completely colour-blind narrative would deliver  $\rho^- \approx 0$ for the negative network.

\begin{figure}[t]
\begin{center}
\includegraphics[width=0.7\columnwidth,angle=0]{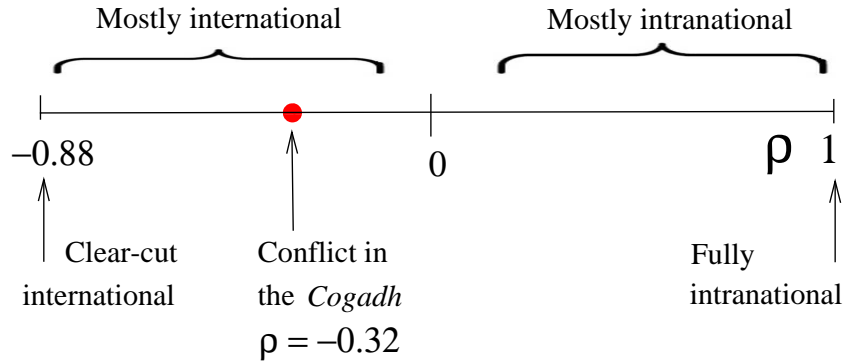}
\caption{Graphical representation of the main conclusion of this paper.
The spectrum of values of {categorical} assortativity for networks of the 
conflictual-{\CG} type ranges from $\rho = -0.88$ to $\rho = 1$.
Negative values of $\rho$ correspond to various degrees of the traditional picture of international hostilities with $\rho = -0.88$ representing a {clear-cut} Irish-versus-Viking conflict.
Positive values correlate with the revisionist picture of mostly {intranational} conflict.
The analysis presented in this paper shows that the {\CG} hostile network delivers a value {$-0.32$} which, although not clear-cut, lies on the traditional side of the spectrum.
}
\label{figure1p2}
\end{center}
\vspace{0.3cm}
\end{figure}

%\todo{Main conclusion of paper}
We find that  {$\rho^-=-0.32(6)$} if all three kinds of node (Irish, Viking and other) are included in the negative network.
This statistic is to be compared to the  theoretical minimum  {${\rho}_{\rm{min}}^-=-0.88(4)$.}
If unassigned nodes are omitted, one finds {$\rho^-=-0.37(6)$} (with ${\rho}_{\rm{min}}^-=-1$).
Thus our measured values for categorical assortativity on the negative (conflictual) networks are themselves negative.
This means that picture of a primarily {intranational} conflict is not supported by data contained in {\CGG}.
However, the conflict is not clear-cut international either; it is a narrative in which the highest proportion of conflict is  presented as being between Irish and Viking but 
with significant amounts of green-on-green and blue-on-blue conflict too.
On the spectrum from international to {intranational} conflict, representing various degrees of the traditional to the revisionist views, 
the negative {\CG} networks are firmly on the traditional side but at a moderate and not a limiting value.
{This spectrum is represented graphically in {{Figure~\ref{figure1p2}}}.}
This is the main conclusion of our paper and is our contribution to the 250-years old debate mentioned in the Introduction.

%...................................................................................
%...................................................................................

%%%%%%%%%%%%%%%%%%%%%%%%%%%%%%%%%%%%%%%%%%%%%%%%%%%%%%%%%%%%%%%%%%%%%%%%%%%%%

\begin{table}[t]
\caption{Categorical assortativities. 
The first column identifies whether all nodes (Irish, Viking and other) are included in the determination of ${\rho}$ or if the unassigned (other) nodes are excluded.
In the former case, ${\rho}_{\rm{min}}$ is determined by Eq.(\ref{rhomin}). 
In the latter case, it is $-1$.
The second column identifies whether all remaining links are included or whether Viking-on-Viking edges are omitted.
}
\begin{center}
\begin{tabular}{|c|l|c|c|}  \hline
Nodes                      & Edges       
																		              & \multicolumn{1}{|c|}{Positive}     
																									               & \multicolumn{1}{c|}{Negative}      
																																                 \\
                             &           
																		              & \multicolumn{1}{c|}{Network}
																											           & \multicolumn{1}{c|}{Network}  
																																                  \\
                             &         
																		              & \multicolumn{1}{c|}{($\rho^+$)}
																									               & \multicolumn{1}{c|}{($\rho^-$)}
																																                     \\
\hline 
        &                                 &              &                \\
\multirow{3}{*}{\rotatebox{90}{{All nodes}}}
\multirow{3}{*}{\rotatebox{90}{{included~}}}
~       & Include all           edges     &  \multicolumn{1}{r|}{0.65(3)}     & \multicolumn{1}{r|}{-0.32(6)}  \\ 
~       & Omit Viking-on-Viking edges     &  \multicolumn{1}{r|}{ }     & \multicolumn{1}{r|}{-0.45(5)}      \\
~       & ${\rho}_{\rm{min}}$             & \multicolumn{1}{r|}{-0.62(3)}     & \multicolumn{1}{r|}{-0.88(4)}      \\
        &                                 &              &                \\
\hline
\multirow{3}{*}{\rotatebox{90}{{Other nodes~}}}
\multirow{3}{*}{\rotatebox{90}{{omitted~~~~}}}
        &                                 &              &                \\
~       & Include all remaining edges     & \multicolumn{1}{r|}{0.72(3)}      & \multicolumn{1}{r|}{-0.37(6)}  \\ 
~       & Omit Viking-on-Viking edges only&\multicolumn{1}{r|}{ }      & \multicolumn{1}{r|}{-0.53(4)}      \\
~       & ${\rho}_{\rm{min}}$             & -1~~~~~~~          & -1~~~~~~~             \\
        &                                 &              &                \\
\hline
\end{tabular}
\end{center}
\label{table5p5}
\end{table}
%%%%%%%%%%%%%%%%%%%%%%%%%%%%%%%%%%%%%%%%%%%%%%%%%%%%%%%%%%%%%%%%%%%%%%%%%%%%%

The assortativity analysis thus far probes the extent to which conflict or harmony reigns within or between the two groups.
However, one may argue that the  revisionist concern is  with the Irish side. 
The claim is that the conflict is primarily within the Irish community --- not that it is both within the Irish cast and within the Viking set. 
Clearly there was a great degree of such conflict too; e.g., Ryan states
%(page 6) 
``The Norse were traditionally unscrupulous in preying upon one another''~\cite{Ryan1938}.
{(See also Ref.\cite{Downham2012a}.)}
Therefore, one may argue that Viking-on-Viking conflicts could contaminate our measurements.
Our aim is to determine whether the Irish are mostly in conflict with other Irish or with Vikings;
in this sense, the fact that the Vikings were also fighting amongst themselves is irrelevant.

To investigate further, we remove all Viking-on-Viking links from the negative sub-network. 
Recalculating the categorical assortativity delivers  {$\rho^- = -0.45(5)$} 
[{$\rho^- = -0.53(4)$} if the unassigned nodes are removed] which indeed is larger in 
magnitude  than the previous measure (the assortative Viking-on-Viking edges having been removed).
But it is still not a clear-cut Irish-versus-Viking picture;
i.e., it is not close to {${\rho}_{\rm{min}}^-=-0.88(4)$} (or $-1$ in the case where unassigned nodes are removed).
Thus our conclusions are unchanged.
These statistics are listed in  {{Table~\ref{table5p5}}}.

In Appendix~\ref{AppBassort}, to overcome the awkwardness of network-dependent $\rho_{\min}$-values, we introduce a renormalised categorical assortativity measure that ranges from $-1$ in the case of fully disassortative networks through zero for uncorrelated networks to $1$ for fully assortative networks. 
We also present in Table~\ref{tableBp1} an alternative to Table~\ref{table5p5}, using these renormalized values.

In summary, we conclude that the character networks embedded in the {\CGG} do not support clear-cut traditionalist or revisionist depictions of the Viking Age in Ireland.
Instead they support a moderate traditionalist picture of conflict which is mostly between Irish and Viking characters, but with significant amounts of hostilities between both sides as well.

%%%%%%%%%%%%%%%%%%%%%%%%%%%%%%%%%%%%%%%%%%%%%%%%%%%
\section{Discussion}
\label{newsec5}
%%%%%%%%%%%%%%%%%%%%%%%%%%%%%%%%%%%%%%%%%%%%%%%%%%%

%\mycheck{\blue{Popular view}} 
%\todo{Context of our study - very old debate}
The popular tradition associated with the Viking Age in Ireland and the events of Clontarf in 1014 is that Brian's principal opponents were Vikings. 
%Ryan  said Page 265: Contemporary annalists “portray Clontarf in terms consistent with the later popular tradition.”
Following Charles O'Connor in 1766,  in  1938 {John} Ryan \cite{Ryan1938} published what has been described as an ``assault'' \cite{IrishTimes} on that traditional interpretation.
Instead of a ``clear-cut'' Irish versus Norse conflict, the revisionist claim is that it was a struggle primarily between Irish forces.
%Page 49 of Ryan: ``Fundamentally, then, the issue at Clontarf was the determination of the Leinstermen to maintain their independence against the High-''
%It has been claimed that this revisionist version of ``nothing more than a bloody local quarrel'' has become  ``the new orthodoxy'' \cite{IrishTimes}. 
With the millennial anniversary of the Battle of Clontarf, 
Se{\'{a}}n Duffy attacked ``the new orthodoxy'' \cite{IrishTimes} and  launched a counter-revisionist defence 
of the traditional picture \cite{Duffy}.
His judicious use of {{\CGG}}  and other texts leads him to conclude that ``The Battle of Clontarf was an international contest'' \cite{IrishTimes}.
This  view has itself come in for criticism \cite{Duffyreview2} 
and the  anniversary reinvigorated lively discussions and healthy debate amongst experts and the wider public.
This and the 150th anniversary of Todd's famous translation \cite{Todd} form the context in which the above results are presented.

{\CGG} is a skillfully written propagandistic text, replete with bias, exaggerating virtues and vices of many of its characters \cite{Todd,NMbias,NMannals}.
It has been used to support arguments from both sides of the debate.
Duffy describes it as a ``long narrative of Irish conflict with the Vikings'' \cite{Duffy}.
%E.g., page 222.
Downham states ``Evidently the conflict was much more than an internal squabble between an Irish over-king and some reluctant subjects'' \cite{Downham2014}.
Etchingham, on the other hand, in reviewing Ref. \cite{Duffy}, stresses that ``even Cogadh actually identifies the Leinstermen as principal rebels''~\cite{Duffyreview2}. 
% Duffyreview2 refers to page 121 of Duffy.
From the side opposing Brian at Clontarf, the {\CG} gives the majority of the slain (3100 out of 5600) as Irish \cite{Todd,Ryan1938}, 
%\todo{Tallies have been used in previous studies}
tallies which could be viewed as supporting the picture of a mostly domestic conflict.
At least these tallies show that {\CGG} does not pretend that Viking slain exceed 
the numbers of Leinstermen in order to ``internationalise'' the story.
This may suggest that, interpolations notwithstanding \cite{DownhamAnn}, even if the {\CG} exaggerated qualities, it may not have exaggerated quantities (at least not by much). 
Indeed, Ryan believes that 
%(page 39) 
the account of the actual battle of Clontarf in the {\CG} is ``incomparably the most reliable''.

%``From the closeness of the contest,'' Ryan (page 28) deduces ``that there was no excessive disparity in strength''  
% between Brian's forces and that of his opponents in the Battle of Clontarf itself. 
% He estimates that the armies on each side contained tens of thousands (page 46) of fighting men. 
% He says (page 38 and again on page 48/49) that the Leinster to Norse ratio was 2:1. 
% Ryan (page 45) states that  CGG gives Brian's losses as 1,600, but  ``Presumably this estimate 
% does not include the freemen of status less than noble who were slain.'' 
% He states (page 45) that, of Brian's opponents, the Irish losses were 3,100 (= 2,000 Leinster proper + 1,100 U{\'{\i}} Chendselaig) 
% and the Norse losses were 2,500.

% Todd says:
% Page cxc: On the side of the Danes there fell-1. Brodar, son of Osll (Floel?) earl of Caer Ebroc or York, "with a thousand plundering Danas, 
% both Saxons and Locblains."
% Page cxci: ``Of the forigners of Dublin were slain 2,000
% ...
% Page cxci: Then followa a Jist of the Irish chiefteins who fell on the Danish aide.
% With these fell 2,000 of the Leinlter men and 1,100 of the Ui Ceinnlelaigh
% the total loss of the enemy being 66,000, which ia no doubt exaggerated.
% 
% See also page 
% 
% 

%\todo{We go beyond tallies}
In  {the above considerations} we have gone beyond a simple tally of the slain and performed a character-network analysis of {\CGG}. 
Since this is wholly independent of the tone of the account (``bombastic'' and ``partisan'') and its shortcomings 
(``telescoping'' of events and ``cavalier'' attitude  to chronology), we considered this approach a judicious use of the text.
To contribute to the debate as to the nature of the Viking Age in Ireland as set down in the {\CG}, 
we applied a  measure of categorical assortativity  which is capable of taking proportions of Irish and Viking nodes into account. 
%\todo{But our method has limitations too}
As we have stressed throughout, any statistical analysis is only as good as the data it draws upon and here all of our data comes directly from the {\CG} text. 
Any conclusions about the implications of our study for the reality of the Viking Age in Ireland have to be made in combination with knowledge from  humanities literature on the topic.
Humanities scholars agree that, to some degree, historical sources lie behind the {\CG}.
But they differ as to their extent.
If, having assessed the evidence, one believes {\CGG}, in the main, to be unreliable, invented or concocted then little can be drawn from our study about reality.
%Even if, rhetoric notwithstanding,  the author of the {\CG} attempted to record as much as possible about encounters and relationships between personages \cyan{of the time,} we have no way of knowing if certain types of interactions are under-reported relative to others. 
%\sout{(e.g., an Irish author may not have been able to record interactions between Vikings with an accuracy comparable to that with which he  recorded interactions involving Irish players).}
Even in this case, however, the text (and hopefully this paper) still delivers information on how medieval writers sought to, or were able to, portray the composition of societies. 
%\red{Although he may have manipulated events and exaggerated virtues and vices of individuals, } 

%\todo{But our method is OK and gives new insights}
A less doubtful assessment of the evidence may offer hope that a reasonable proportion of characters and their interactions reflect the reality of the age (and we have seen that our network statistics are robust; even omitting Viking-Viking interactions does not alter the broad conclusions of our study).
Indeed, since the {\CG} author scarcely anticipated a complexity-scientific analysis nearly 1000 years thenceforth, one might expect the networks to be less encumbered by the bias and partisanship that permeates more qualitative aspects of the text.
In this sense, the networks approach delivers unique insights in that it extracts a perhaps unintended message from his time, namely new, quantitative  knowledge of  the Viking Age in Ireland.

%%%%%%%%%%%%%%%%%%%%%%%%%%%%%%%%%%%%%%%%%%%%%%%%%%%
\section{{Conclusions}}
\label{sec6}
%%%%%%%%%%%%%%%%%%%%%%%%%%%%%%%%%%%%%%%%%%%%%%%%%%%

%\todo{Main conclusions}
The purpose of this paper is to gain quantitative insight into the complexity and conflicts of the Viking Age in Ireland as described in \CGG.
A literal interpretation of ``the popular tradition of Clontarf as wholly an Irish-Norse'' conflict \cite{Ryan1938} 
would suggest a strongly negative value of categorical assortativity for the negative (conflictual) network.
On the other hand, the revisionist picture of a ``civil war'' \cite{OConnor1766}, an ``internal struggle''  \cite{Corrain72}, with Leinster as the ``predominant element'' \cite{Ryan1938} or ``principal rebels'' \cite{Duffyreview2}, suggests a positive value of categorical assortativity for the negative network.
The primary outcome of our investigation is our measured value of the associated metric and we find a negative value, supportive of the traditional picture.
But its magnitude is moderate, suggesting  that, at least in network terms,  {\CGG}  does not describe a fully ``clear-cut'' Irish versus Norse conflict.
The power of our analysis is that we can quantify this statement and the 
value $\rho = -0.32$ means {\CGG}  describes the Viking Age in Ireland as predominantly an Irish-Norse conflict but it is not wholly so.

%\todo{Future work}
There are a number of other ways in which this work can be extended.
Like Refs.\cite{EPL,EPJB}, the present analysis is based on static networks. 
These freeze the narrative progress and capture  the plot ``all at one glance in a visual display of its character network'' \cite{Rhody}.
Static networks are particularly advantageous for {\CGG}  which, although believed to have been composed following some of the annals, paid limited regard to chronology {\cite{NMannals, DownhamAnn}.}
Nonetheless, dynamical properties are also of interest and should be investigated in the future \cite{ACS}.
It would be interesting to see if temporal networks can help restore some of the chronology to {\CGG} \cite{McCarthy2008}.
Directed and weighted networks also offer obvious routes for wider study.
Furthermore, motivated by the Ua Ruairc example, it would also be interesting to investigate if the structural imbalance in some network triads could be developed to give a way to spot other  potential interpolations, not least because the survival of only one complete manuscript limits opportunities to identify interpolations through comparisons \cite{DownhamAnn}.
Another question is how the {\CG} narrative compares to others of the epic genre \cite{EPL,EPJB,Ossian}.
A comparison to the {\emph{Iliad}} would be especially important as a link to  an Irish account of the Trojan War ({\emph{Togail Tro{\'{\i}}}} --- ``The Destruction of Troy'') has been suggested before by humanities scholars, using traditional methods~\cite{Bugge05,Goedheer,NMdating,MhaonaighTroy}.
% Bugge05 page XVII says ``The author of Cogadh Gaedhel must have known the tale called »The Destruction of Troy« [Togal Troi). This Irish version of the Argonautic expedition and the Destruction of Troy is preserved in the Book of Lcinster, in a transcript from about the middle of the 12th century (and''
It would be interesting to continue such comparative  investigations at a more detailed level in future studies.
% A plethora of other quantitative approaches and suggestions are contained in the compendium \cite{MMM}.

A criticism sometimes leveled at the character-network approach is that it brings little new; merely confirming knowledge already gained from traditional approaches to humanities. 
The rebuttal to such criticism is that agreement is precisely what one would expect from a new approach which is valid and still evolving.
The quantitative determination of categorical assortativity in this paper, and its precise placement of {\CGG} along the spectrum from the international to the {intranational}, is a new development in the evolution of this field. 
In that sense, our paper goes beyond limitations identified in some previous works in that it generates a new quantitative element to an unfinished debate in the humanities.

%%%%%%%%%%%%%%%%%%%%%%%%%%%%%%%%%%%%%%%%%%%%%%%%%%%%%%%%%%%%%%%%%%%%%%%%%%%%%%%%%%%%%%
\vspace{0.2cm}
{\section*{Ethics Statement}}
%%%%%%%%%%%%%%%%%%%%%%%%%%%%%%%%%%%%%%%%%%%%%%%%%%%%%%%%%%%%%%%%%%%%%%%%%%%%%%%%%%%%%%
\noindent
An ethics statement does not apply to this manuscript.

%%%%%%%%%%%%%%%%%%%%%%%%%%%%%%%%%%%%%%%%%%%%%%%%%%%%%%%%%%%%%%%%%%%%%%%%%%%%%%%%%%%%%%
\vspace{0.2cm}
{\section*{Data Accessibility}}
%%%%%%%%%%%%%%%%%%%%%%%%%%%%%%%%%%%%%%%%%%%%%%%%%%%%%%%%%%%%%%%%%%%%%%%%%%%%%%%%%%%%%%
\noindent
Data are available from  
{\url{https://github.com/ralphkenna/CGG}}.

%%%%%%%%%%%%%%%%%%%%%%%%%%%%%%%%%%%%%%%%%%%%%%%%%%%%%%%%%%%%%%%%%%%%%%%%%%%%%%%%%%%%%%
\vspace{0.2cm}
\section*{Competing Interests}
%%%%%%%%%%%%%%%%%%%%%%%%%%%%%%%%%%%%%%%%%%%%%%%%%%%%%%%%%%%%%%%%%%%%%%%%%%%%%%%%%%%%%%
\noindent
We have no competing interests.

%%%%%%%%%%%%%%%%%%%%%%%%%%%%%%%%%%%%%%%%%%%%%%%%%%%%%%%%%%%%%%%%%%%%%%%%%%%%%%%%%%%%%%
\vspace{0.2cm}
\section*{Author Contributions}
%%%%%%%%%%%%%%%%%%%%%%%%%%%%%%%%%%%%%%%%%%%%%%%%%%%%%%%%%%%%%%%%%%%%%%%%%%%%%%%%%%%%%%
\noindent
JY harvested the data and performed the statistical analysis;
RK conceived of the study, coordinated the study and drafted the manuscript;
MMC harvested the data;
PMC performed the statistical analysis;
All authors discussed and interpreted the results and gave final approval for publication.

%%%%%%%%%%%%%%%%%%%%%%%%%%%%%%%%%%%%%%%%%%%%%%%%%%%
\vspace{0.2cm}
\section*{Acknowledgments}
%%%%%%%%%%%%%%%%%%%%%%%%%%%%%%%%%%%%%%%%%%%%%%%%%%%
\noindent
We would like to thank {Denis Casey,}
M{\'{a}}ire N{\'{\i}} Mhaonaigh and Justin Tonra
for carefully reading the manuscript and for helpful comments. 
We thank Thierry Platini for regular discussions about network science.

%%%%%%%%%%%%%%%%%%%%%%%%%%%%%%%%%%%%%%%%%%%%%%%%%%%%%%%%%%%%%%%%%%%%%%%%%%%%%%%%%%%%%%
\vspace{0.2cm}
\section*{Funding statement}
%%%%%%%%%%%%%%%%%%%%%%%%%%%%%%%%%%%%%%%%%%%%%%%%%%%%%%%%%%%%%%%%%%%%%%%%%%%%%%%%%%%%%%
\noindent
JY was supported by the Doctoral College for the Statistical Physics of Complex Systems. 
{JY and RK were supported by} the EU Marie Curie IRSES Network PIRSES-GA-2013-612707
DIONICOS - Dynamics of and in Complex Systems
funded by the European Commission within the FP7-PEOPLE-2013-IRSES Programme (2014-2018).
{PMC was funded by a European Research Council Advanced Investigator grant  to Robin Dunbar (Grant No. 295663).}

\newpage

%%%%%%%%%%%%%%%%%%%%%%%%%%%%%%%%%%%%%%%%%%%%%%%%%%%%%%%%%%%%%%%%%%%%%%%%%%%%%%%%%%%%%%%%%%%%%%%%%%%%%%%%%%%%%%%%%%%
\appendix

%%%%%%%%%%%%%%%%%%%%%%%%%%%%%%%%%%%%%%%%%%%%%%%%%%%%%%%%%%%%%%%%%%%%%%%%%%%%%%%%%%%%%%%%%%%%%%%%%%%%%%%%%%%%%%%%%%

%%%%%%%%%%%%%%%%%%%%%%%%%%%%%%%%%%%%%%%%%%%%%%%%%%%%%%%%%%%%%%%%%%%%%%%%%%%%%%%%%%%%%%%%%%%%%%%%%%%%%%%%%%%%%%%%
\section{Ireland during the Viking Age}
\label{AppendixA}
%%%%%%%%%%%%%%%%%%%%%%%%%%%%%%%%%%%%%%%%%%%%%%%%%%%%%%%%%%%%%%%%%%%%%%%%%%%%%%%%%%%%%%%%%%%%%%%%%%%%%%%%%%%%%%%%%%%

%..................................................................................
\begin{figure}[t]
\centerline{\includegraphics[width=0.3\textwidth]{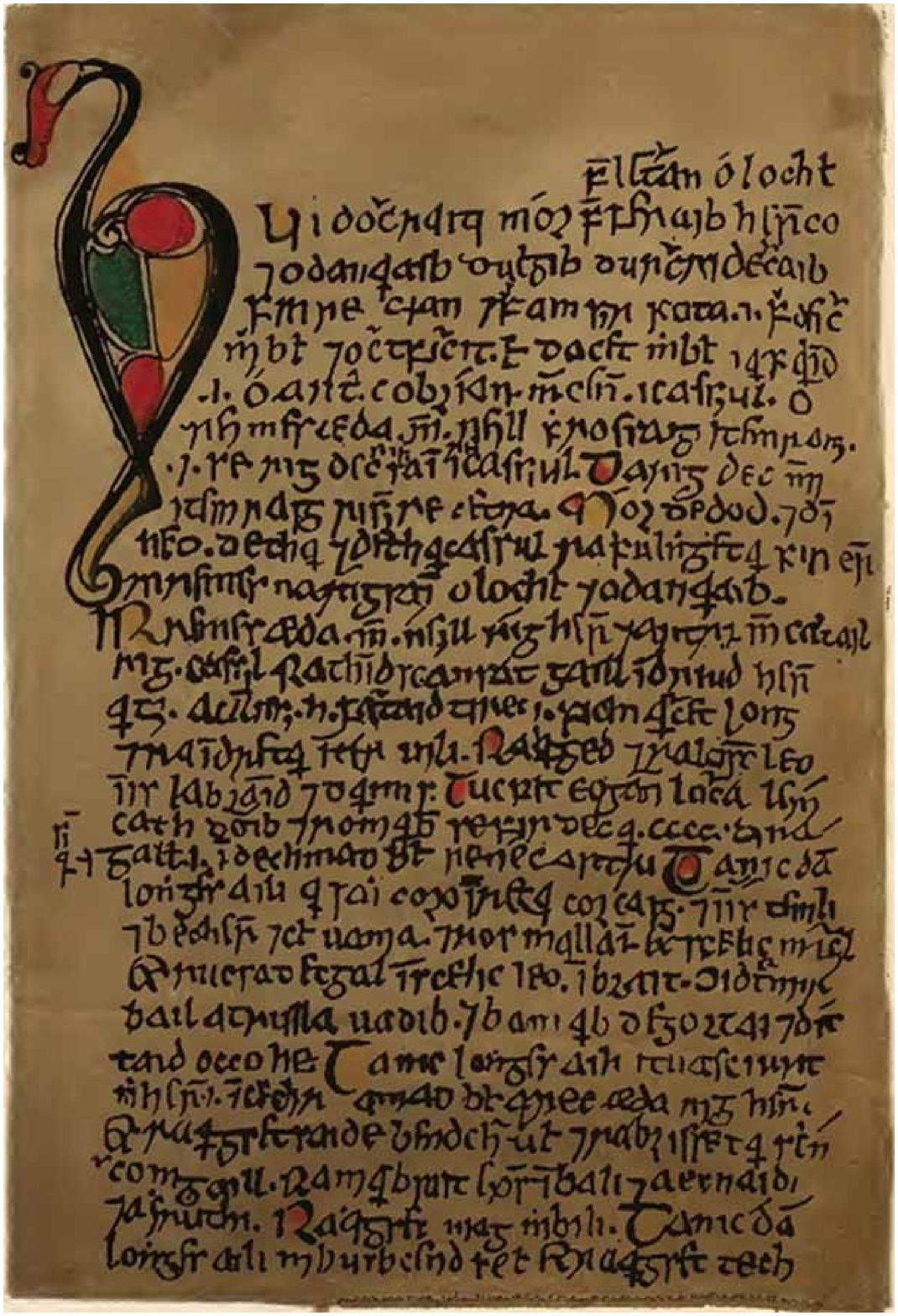}~~~~~~~~~~~~~~~\includegraphics[width=0.35\textwidth]{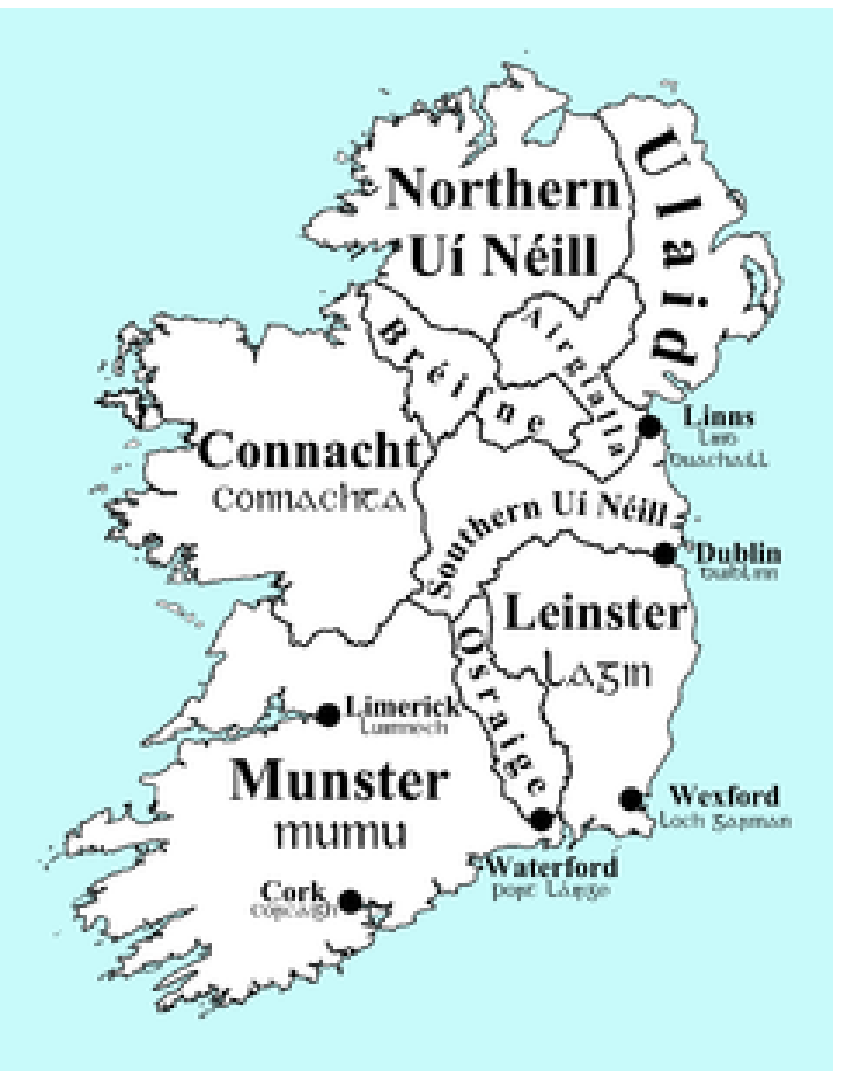}}
\caption{Left: {Image of the nineteenth-century facsimile of the opening page of {\CGG} which was reproduced in Todd's edition \cite{Todd}. }
% OR  ``Reproduction of the Cogadh Gaedhel re Gallaibh manuscript in James H. Todd's edition'' which is how it is put in 
% https://dh.tcd.ie/clontarf/%20%27Cogadh%20Gaedhel%20re%20Gallaibh%27%3A%20The%20War%20of%20the%20Irish%20with%20the%20Foreigners
%The opening page of the oldest extant copy of {\CGG} from the twelfth century Book of Leinster.
Right: The main kingdoms of Ireland {circa 900AD with principal (Viking) towns.} 
}  
\label{figureAp1}
\end{figure}
%...................................................................................

The five provinces referred to in the main text are Connacht, Leinster, Ulster, Meath and Munster.
{Their names are associated, respectively, with member populations called the Connachta, the Laigin, the Ulaid, and the kingdoms of Mide and of Mumu.
%\blue{\sout{These five provinces were called called {\emph{c{\'{o}}iceda}} or fifths.}}
%Medieval Connacht and Munster roughly correspond to today's provinces of the same names.
The modern province of Ulster  encompasses the territories of the Northern U{\'{\i}}~Neill and Ulaid (from which Ulster derives its name), as well as parts of Br{\'{e}}ifne and Airg{\'{\i}}alla.}  
Mide, associated with the Southern U{\'{\i}} N{\'{e}}ill,  mainly comprised the modern county Westmeath and part of Meath and has been subsumed into the modern Leinster.
% DC said Medieval Ulaid was mainly located to the east of the Bann and Bush rivers i.e. essentially the modern counties of Antrim and Down.
%DC said The Northern Uí Néill were not members of the Ulaid (and would have been indignant at the notion of calling their territory Ulaid/Ulster).  I generally use 'the north of Ireland' when writing about this wider geographical area as it avoids conflicts in terminology}
%{\emph{C{\'{u}}ige}} (literally ``fifth part'') remains the modern Irish word for ``province''.
In the tenth century the main rivalry for claims to high kingship
% I got the term ``ceremonial'' from http://www.rootsweb.ancestry.com/~irlkik/ihm/uinkings.htm
 of Ireland was between the northern and southern branches of the U{\'{\i}} N{\'{e}}ill. 
Their dominance was ended by Brian Boru.

In the ninth century, Cork, Dublin, Limerick, Waterford and Wexford all developed from Vikings base camps to more permanent settlements.
See Figure~\ref{figureAp1} which, alongside an image {adopted from the Book of Leinster for Todd's edition} of {\CGG},  includes a map outlining the political structure of Ireland {about 900AD}.

%%%%%%%%%%%%%%%%%%%%%%%%%%%%%%%%%%%%%%%%%%%%%%%%%%%%%%%%%%%%%%%%%%%%%%%%%%%%%%%%%%%%%%%%%%%%%%%%%%%%%%%%%%%%%%%%%%%
\section{Scalar and categorical assortativity}
\label{AppBassort}
%%%%%%%%%%%%%%%%%%%%%%%%%%%%%%%%%%%%%%%%%%%%%%%%%%%%%%%%%%%%%%%%%%%%%%%%%%%%%%%%%%%%%%%%%%%%%%%%%%%%%%%%%%%%%%%%%%%

In the main text we used two different forms of assortativity: the degree assortativity $r$ and the measure $\rho$. 
The first of these is an example of {\emph{scalar assortativity}} --- it quantifies the tendency of nodes whose degrees have similar values to associate with each other.
In determining $r$, it is important to account for nodes possibly having similar but not identical values; e.g., high degree nodes may tend to mix with other high degree nodes without them having to have precisely the same $k$-values.
The second is {\emph{categorical}} --- it measures tendencies for nodes belonging to the same category to link to each other.
In the categorical case, two nodes either have the same  attributes or they do not; there is no question of degrees of similarity here.
Therefore we require two different formulae to quantify  scalar and categorical assortativity.

Scalar assortativity is simply given by Pearson's correlation coefficient, i.e., it is the covariance of two variables normalised by the product of their standard deviations. 
The normalization factor ensures that the assortativity takes values in the range $[-1,1]$.
Networks with a degree value $r>0$ are termed {\emph{degree assortative}}.
If the measured value of $r$ is negative, the network is deemed {\emph{degree disassortative}}.
Since the theoretical bounds on scalar assortativity are the same for all networks, comparisons of assortativity between them are straightforward and meaningful.

Many networks tend to evolve towards their maximum-entropy state unless otherwise constrained \cite{Bianconi07}. 
Such maximum-entropy states are usually disassortative because disassortative configurations are more abundant than assortative ones \cite{Johnson2010}.
For this reason, non-social networks are usually degree-disassortative.
Social networks, on the other hand,  are usually uncorrelated or assortative. 
This can be explained by homophily; highly connected people tend to link together \cite{Newman2002,Newman2003}.
The lack of disassortativity in the positive networks, as seen in Table~{\ref{table5p2}}, is a common feature of epic narratives.
It is a signal of the presence of a non-trivial social or narrative force --- driving them away from their maximum-entropy, anticorrelated (disassortative) states. 
In this sense, positive character networks, including those of {\CGG} are more like social networks than unlike them.

For categorical assortativity, consider the nodes $i$ of a network having attributes $c_i$ which could be colours (e.g., green,  blue or grey) as in the main text.
We require the difference between the fraction of edges that exist between nodes of the same attribute and the fraction of such edges we would expect if the nodes were connected at random regardless of the nodes' attributes (i.e., if the linking process were ``colour blind'').
It is defined as follows~\cite{Newman2002,Newman2003}.

The total degree of the network is $\sum_{i=1}^N{k_i} = 2M$ (twice the number of edges because each edge is double-counted).
Let $c$ and $c^\prime$ denote categorical variables and let $e_{cc^\prime}$ denote the density of directed edges in the network pointing from nodes of type $c$ to nodes of type  $c^\prime$.
We note that  $e_{cc^\prime} =  e_{c^\prime c}$ if the network is undirected.
We define the density of degrees associated with nodes of type $c$ as
\begin{equation}
 a_{c} =  \sum_{c^\prime}{ e_{cc^\prime}} = \frac{1}{2M}\sum_i{k_i \delta_{c_i c}},
 \label{M5}
\end{equation}
and have the sum rule
\begin{equation}
 \sum_{c}{ a_{c}} = \sum_{cc^\prime}{ e_{cc^\prime}} = 1.
\label{sumrule}
\end{equation}
The  {\emph{modularity}} is defined as 
\begin{equation}
 Q
 = 
 \sum_c{\left({e_{cc} - a_c^2}\right)} .
\label{Q}
\end{equation}
The categorical assortativity $\rho$ is obtained by normalising the modularity so that its maximum value is 1 (as is the case for the scalar assortativity).
If the network is fully assortative, {\emph{all}} edges connect nodes of the same type.
Therefore the  normalising factor for $Q$ is given by Eq.(\ref{Q}) with ${\sum_c{e_{cc}}}$ set to 1.
This motivates the definition
\begin{equation}
 \rho = \frac{ \sum_c{\left({e_{cc} - a_c^2}\right)} }
                             { 1 - \sum_c{a_c^2} }.
\label{rho}
\end{equation}
The minimum possible value of this quantity is obtained when all edges connect nodes of different types ($e_{cc}=0$ for all $c$) and is
\begin{equation}
 \rho_{\rm{min}} = \frac{ -\sum_c{ a_c^2 }}
                             { 1 - \sum_c{a_c^2} }.
\label{rhomin}
\end{equation}

Fully disassortative, undirected networks with only two categories have $\rho_{\rm{min}}=-1$.
However, the minimum value for $\rho$  is not generally $-1$ if more categories are involved.
While the absence of assortativity means that $\sum_c{e_{cc}} = 0$ for any number of categories, the lack of diretedness that assures the symmetry between the categories only happens when there are two of them.
This property, together with  Eq.(\ref{sumrule})  trivially gives $\rho=-1$.
More generally, $ \rho_{\rm{min}} $ lies between $-1$ and $1$.

The reason why $\rho_{\rm{min}}$ is not $-1$ is for a perfectly disassortative network is that such a network more closely resembles a random network than does a perfectly assortative one when there is more than two categories.
I.e., random mixing mostly mixes unlike nodes and disassortativity does the same. But assortativity mixes like nodes. This is why the minimum value of $\rho$ is closer to the value for a random network $\rho=0$ than is the maximum value $\rho=1$.
In the main text, we have to be mindful of this when interpreting the categorical assortativity for the negative network.
However, we could easily introduce a measure which is $-1$ for a fully disassortative network as follows.

The modularity in Eq.(\ref{rho}) is defined with respect to the expected density of edges between nodes of the {\emph{same}}  category if the network were assembled without regard to category. 
This was appropriate for the measurement of {\emph{assortativity}}.
To directly measure {\emph{disassortativity}} instead, we focus on edges between node of {\emph{different}} categories and introduce
\begin{equation}
\bar{\rho} = - \frac{ \sum_{c,c^\prime}\nolimits'{\left({e_{cc^\prime} - a_ca_{c^\prime}}\right)} }
                             { 1 - \sum^\prime_{c,c^\prime}a_ca_{c^\prime} },
\label{rhobar}
\end{equation}
where the prime on the summation means that it is taken over $c$ and $c^\prime$ values such that $c\ne c^\prime$
and the leading minus sign is to ensure that disassortative networks have negative $\bar{\rho}$-values, in line with their negative $\rho$-values.

Eq.(\ref{sumrule})  gives 
\[
 \sum_{c,c^\prime}\nolimits'{e_{cc^\prime}} = 1-\sum_c{e_{cc}},
\]
enabling us to write
\begin{equation}
\bar{\rho} =  \rho \left({\frac{1}{\sum_c{a_c^2}}-1}\right).
\label{rhobarrho}
\end{equation}
From Eq.(\ref{rhomin}), this may be written 
\begin{equation}
 \bar{\rho} = - \frac{\rho}{\rho_{\rm{min}}}.
\label{rhobarnorm}
\end{equation}
In other words, $\bar{\rho}$ is simply the assortativity normalised by its minimum possible value (which is negative).
This has the advantage that its value is $1$ for a fully disassortative network; however a fully assortative network may have a value of  $\bar{\rho}$ which exceeds 1.

\begin{table}[t]
\caption{The set of renormalised categorical assortativity values $\hat{\rho}$ from Eq.(\ref{rhobar}) presented here is an alternative to Table~\ref{table5p5}. Fully disassortative, uncorrelated,  and assortative networks have  $\hat{\rho}=-1$, $\hat{\rho}=0$  and $\hat{\rho}=1$, respectively. }
\begin{center}
\begin{tabular}{|c|l|c|c|}  \hline
Nodes                      & Edges       
																		              & \multicolumn{1}{|c|}{Positive}     
																									               & \multicolumn{1}{c|}{Negative}       \\
                             &         
																		              & \multicolumn{1}{c|}{Network}
																											           & \multicolumn{1}{c|}{Network}   \\
                             &       
																		              & \multicolumn{1}{c|}{($\hat{\rho}^+$)}
																									               & \multicolumn{1}{c|}{($\hat{\rho}^-$)}\\
\hline 
        &               &      &         \\
\multirow{3}{*}{\rotatebox{90}{{All nodes}}}
\multirow{3}{*}{\rotatebox{90}{{included~}}}
~       & Include all           edges         &  \multicolumn{1}{r|}{0.65(3)}     & \multicolumn{1}{r|}{-0.32(6)}       \\ 
~       & Omit Viking-on-Viking edges         &  \multicolumn{1}{r|}{ }     & \multicolumn{1}{r|}{-0.43(5)}       \\
~       & ${\rho}_{\rm{min}}$                 & -1~~~~~~~          & -1~~~~~~~             \\
        &                                     &              &                \\
\hline
\multirow{3}{*}{\rotatebox{90}{{Other nodes~}}}
\multirow{3}{*}{\rotatebox{90}{{omitted~~~~}}}
        &                                     &              &                \\
~       & Include all remaining edges         & \multicolumn{1}{r|}{0.72(3)}      & \multicolumn{1}{r|}{-0.33(6)}       \\ 
~       & Omit Viking-on-Viking edges only    & \multicolumn{1}{r|}{        }     & \multicolumn{1}{r|}{-0.44(5)}       \\
~       & ${\rho}_{\rm{min}}$                 & -1~~~~~~~          & -1~~~~~~~             \\
        &                                     &              &                \\
\hline
\end{tabular}
\end{center}
\label{tableBp1}
\end{table}

We  therefore introduce a renormalised version of the categorical assortativity that is suitable for all circumstances:
\begin{equation}
 \hat{\rho} = 
  \left\{{ 
	       \begin{array}{ll}
	                                                \rho & \mbox{if $\rho >0$,} \\
                          - \frac{\rho}{\rho_{\rm{min}}} & \mbox{if $\rho <0$.}
	       \end{array}
	}\right.
	\label{rhohat}
\end{equation}
This measure has the desired features that it vanishes in the case of colour-blindness, and it is 1 and $-1$ for fully assortative and fully disassortative networks, respectively.
In Table~\ref{tableBp1}, we list the values of $\hat{\rho}$ for the various networks. 
This may be considered as a renormalised version of Table~\ref{table5p5} of the main text.
The differences between the values entered in the two tables are very small.

% % % % % % % % % % % % % % % % % % % % % % % % % % % 
\section{Network robustness and importance of individual characters}
\label{AppDnames}
% % % % % % % % % % % % % % % % % % % % % % % % % % % 

\begin{table}[t]
\caption{The most important characters of {\CGG} ranked according to their degree, betweenness centrality, closeness, and eigenvector centrality.
%\blue{(We mostly use the spellings of the Index of Todd's version \cite{Todd}.)}
}
\begin{center}
\resizebox{\textwidth}{!}{%
\begin{tabular}{|l|l|l|l|l|l|} \hline  
             & {Rank}   & {Degree}         & {Betweenness  }    &  {Closeness }        &  {Eigenvector}           \\
						\hline
             &  1     & Brian (105)        & Brian (0.42)       & Brian (0.44)       & Brian (0.53)           \\ 
						\multirow{4}{*}{\rotatebox{90}{Unsigned}}    
             &  2     & Sitriuc (62)       & Sitriuc (0.21)     & Sitriuc (0.41)     & Maelmordha (0.28)        \\
             &  3     & Maelmordha (42)    & Ottir (0.16)       & Ottir (0.39)       & {M{\'{a}}el Sechnaill} (0.22)       \\
             &  4     & Ottir (40)         & Aedh Finnliath (0.13)& Gormflaith (0.38)& {Sitriuc} (0.21)         \\
             &  5     & {M{\'{a}}el Sechnaill} (36)    & Ossill (0.11)      & Maelmordha (0.38)  & Gormflaith (0.21)    \\
\hline 
            &  1      & Brian (53)         & Brian (0.28)       & Sitriuc (0.34)     & Brian (0.48)            \\ 
						\multirow{4}{*}{\rotatebox{90}{Positive}} 
            &  2      & Sitriuc (40)       & Sitriuc (0.17)     &  Brian (0.34)		   & Murchadh (0.30)            \\
            &  3      & Maelmordha (38)    & {M{\'{a}}el Sechnaill} (0.11)  & Gormflaith (0.34)  & Maelmordha (0.26)       \\
            &  4      & Gormflaith (34)    & Ottir (0.10)       & Maelmordha (0.32)  & {M{\'{a}}el Sechnaill} (0.26)          \\
            &  5      & Ottir (32)         & Gormflaith (0.10)  & {M{\'{a}}el Sechnaill} (0.32)  & Conaing (0.23)    \\
\hline 
            &  1      & Brian (63)         & Brian (0.63)		    & Brian (0.44) 		   & Brian (0.66)   \\ 
						\multirow{4}{*}{\rotatebox{90}{Negative}}
            &  2      & Sitriuc (25)       & Ottir (0.23)       & {M{\'{a}}el Sechnaill} (0.35)  & Maelmordha (0.23)         \\
            &  3      & Mathgamhain (17)   & Sitriuc (0.23)     & Sitriuc (0.34) 	   & Brodar (Brodir)  (0.22)         \\
            &  4      & Cathal (14)        & Aedh Finnliath (0.16)  & Ottir (0.33)   & {M{\'{a}}el Sechnaill} (0.17)        \\
            &  5      & Olaf Cuaran (12)   & Olaf Cuaran (0.12) & Ivar (0.32)        & Ivar (0.17)             \\
		      	
\hline
\end{tabular}}
\end{center}
\label{tableDp1}
\end{table}

Having investigated the giant component in the main text, we may ask how reliant its integrity is on the  most important characters.
This is a question of robustness and one investigates it by determining the effects of systematic and random removal of nodes or edges.
In the former approach, we remove the most important  nodes one-by-one and monitor how the giant component reduces in size. 
We can then compare this to the results of the latter approach, in which removal of nodes is a random process.

There are a number of ways in which we can decide which are the most important or influential nodes.
One way is to consider that those with highest degree are most important and to remove them first.
Another possibility is to consider nodes with the highest {\emph{betweenness centralities}} \cite{{Freeman}}.
This counts the number of shortest paths (geodesics) which pass through each node \cite{Freeman}. 
To define it, we first write the number of geodesics  between nodes $i$ and $j$ as $\sigma(i, j)$.
We denote the number of these which pass through node $l$ as $\sigma_l(i, j)$.
The betweenness centrality of vertex $l$ is then defined as 
\begin{equation}
g_l = \frac{2}{(N-1)(N-2)}\sum_{i\neq j} \frac{\sigma_l(i,j)}{\sigma(i,j)}.
\label{eqn:betweenness}
\end{equation}
If $g_l=1$, all geodesics pass through node $l$.
If $i,j$ and $l$ represent edges rather than nodes, Eq.~(\ref{eqn:betweenness}) can be interpreted as the {\emph{edge betweenness centrality}} instead.

Other measures of importance include nodes' {\emph{closeness}} and {\emph{eigenvector centralities}}.
The sum of the distances of a given node from all other nodes in a connected graph or component is termed its {\emph{farness}}. 
The reciprocal of farness is a measure of how central a node is and is termed its {\emph{closeness}} \cite{Newman_book}.
{\emph{Eigenvector centrality}} characterises node importance in terms of centralities of its neighbours;   
nodes are deemed influential according to how they are linked to other important nodes \cite{Newman_book}.
Eigenvector centrality is a variant of the ``pagerank'' score used to rank websites.
The leading characters of {\CGG} are listed in {{Table~\ref{tableDp1}}}, ranked according to four different measures: degree; betweenness; closeness and eigenvector centrality.

%...................................................................................
\begin{figure}[t]
\begin{center}
\includegraphics[width=0.41\textwidth,valign=t]{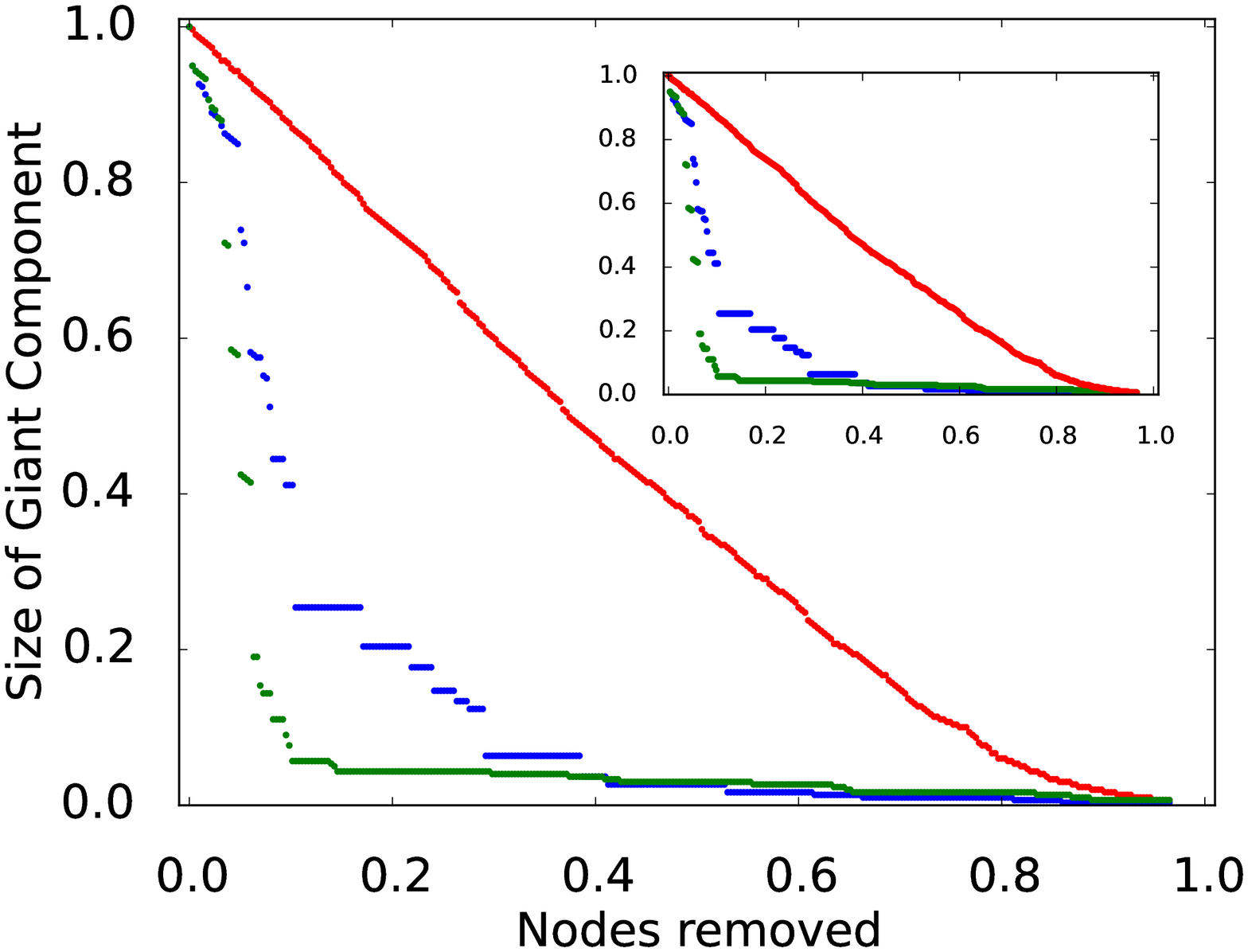}
\includegraphics[width=0.41\textwidth,valign=t]{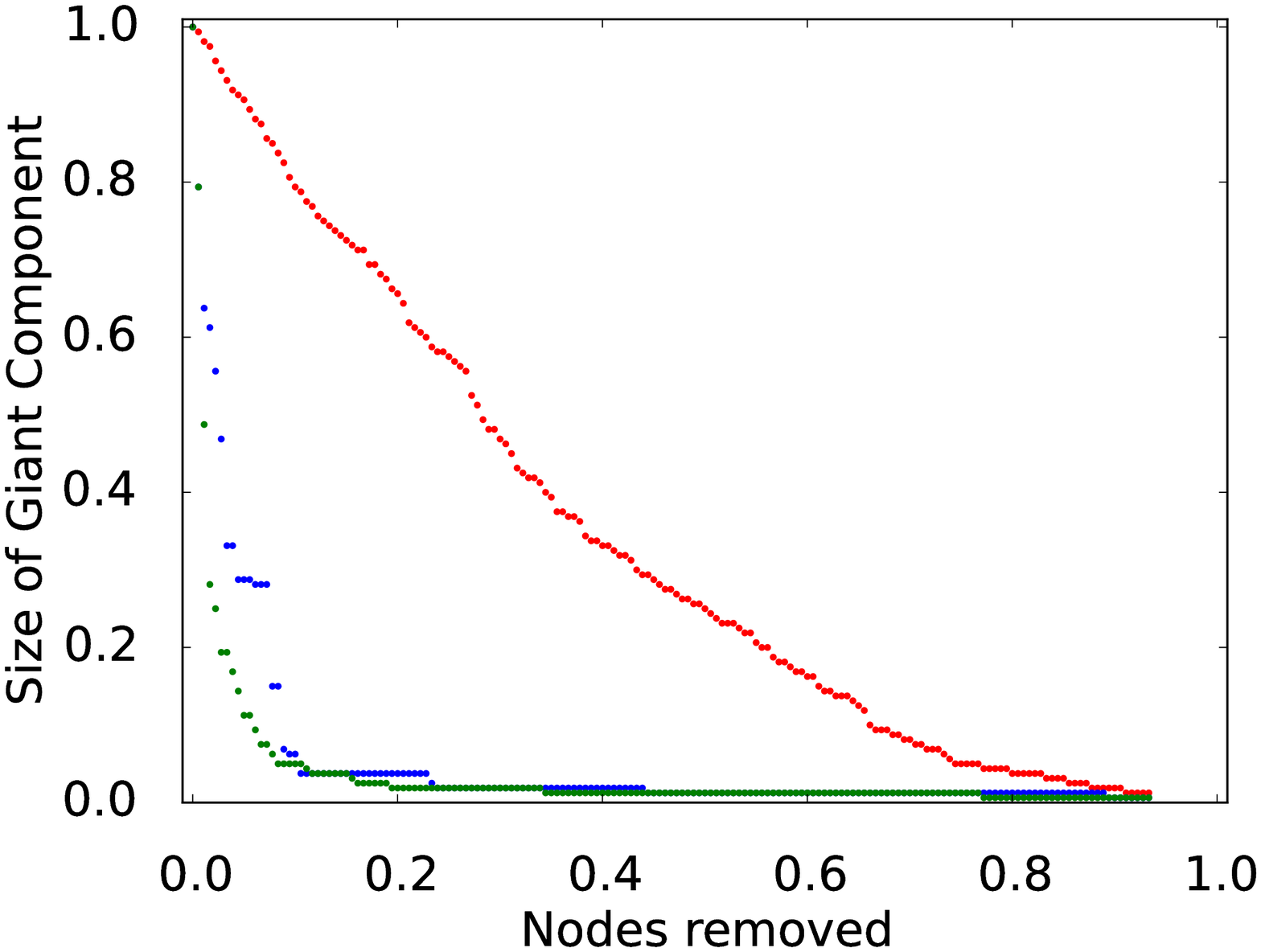}
\includegraphics[width=0.15\textwidth,valign=t]{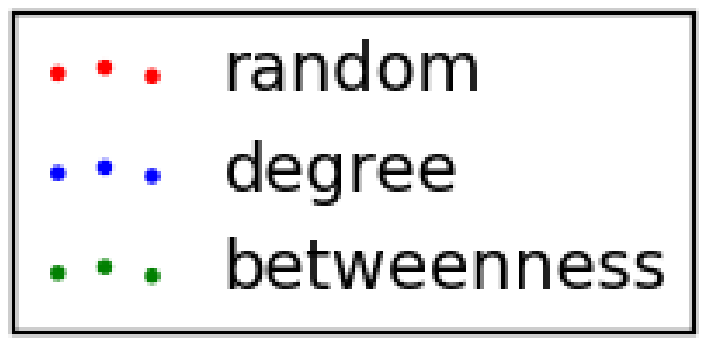}
\caption{The relative sizes of the giant components  as a function of the percentage of nodes removed.
In the left panel the size of the giant component for the unsigned network is given. 
That of the positive network, which has a very similar decay, is given as an insert.
The right panel shows the decay of the giant component of the negative network as nodes are removed. 
The red data points correspond to random removal of nodes  and the blue and green data concern removal by highest degree and betweenness, respectively.
}
\label{figureDp1}
\end{center}
\end{figure}
%...................................................................................

We present the study of robustness for the networks underlying {\CGG}  in {{Figure~\ref{figureDp1}}}.
The main left panel depicts the relative sizes of the giant component of the unsigned network as nodes are removed randomly (red data points), by highest betweenness (blue) and by degree (green).
A similar behaviour is observed for the positive network, shown in the insert. 
The counterpart information for the negative sub-network is contained in the next  panel.
We see that random removal of nodes only has a relatively gradual effect on the giant-component size in all three networks.
Removal by betweenness or by degree has far more devastating consequences.
Removal by betweenness is particularly damaging for the integrity of the full and positive networks whereas, for the negative network, removal by betweenness and degree are about equally effective. 
Details of the effects of node-removal on the relative sizes of the giant components are given in Table~\ref{tableDp2}.

\begin{table}[t]
\caption{The effects of removing the most important characters or of removing characters at random. The entries in the table give the relative size of the giant component after removal of the top 10\% of characters systematically and randomly; the top five characters; and after removal of the most important character, namely Brian Boru.}
\begin{center}
%\resizebox{\textwidth}{!}{%
\begin{tabular}{|l|r|r|r|r|r|r|}
 \hline  
             & \multicolumn{1}{l|}{Remove}  & \multicolumn{1}{l|}{Remove}          & \multicolumn{1}{l|}{Remove}      & \multicolumn{1}{l|}{Remove}   & \multicolumn{1}{l|}{Remove}       & \multicolumn{1}{l|}{Remove}   \\
             & \multicolumn{1}{l|}{10\% by} & \multicolumn{1}{l|}{10\% by}         & \multicolumn{1}{l|}{10\%  }      & \multicolumn{1}{l|}{top 5 by} & \multicolumn{1}{l|}{top 5 by}     & \multicolumn{1}{l|}{Brian}      \\
             & \multicolumn{1}{l|}{degree}  & \multicolumn{1}{l|}{betweenness}     & \multicolumn{1}{l|}{randomly}    & \multicolumn{1}{l|}{degree}   & \multicolumn{1}{l|}{betweenness}  & \multicolumn{1}{l|}{Boru}        \\
\hline 
Unsigned     &  43\%  &  6\%             &  92\%       &  90\%    &   91\%         &  92\%  \\
Positive	   &  47\%  &  7\%             &  83\%       &  85\%    &   85\%         &  86\% \\
Negative     &   6\%  &  5\%             &  81\%       &  69\%    &   58\%         &  85\%  \\
\hline
\end{tabular}
%}
\end{center}
\label{tableDp2}
\end{table}

\clearpage

%
%%%%%%%%%%%%%%%%%%%%%%%%%%%%%%%%%%%%%%%%%%%%%%%%%%%%%%%%%%%%%%%%%%%

\end{document}